\documentclass[12pt]{article}
\usepackage{amsfonts}
\usepackage{epsfig}
\usepackage{graphics}
\usepackage{amsmath}

\begin{document}

\title{{\Large Dirac equation in magnetic-solenoid field}}
\author{S.P. Gavrilov\thanks{%
Dept. Fisica e Quimica, UNESP, Campus de Guaratingueta, Brazil; On leave
from Tomsk State Pedagogical University, 634041 Russia; e-mail:
gavrilovsp@hotmail.com}, D.M. Gitman\thanks{%
e-mail: gitman@dfn.if.usp.br}, and A.A. Smirnov\thanks{%
e-mail: saa@dfn.if.usp.br}}
\date{\today }
\maketitle

\begin{abstract}
We consider the Dirac equation in the magnetic-solenoid field (the field of
a solenoid and a collinear uniform magnetic field). For the case of
Aharonov-Bohm solenoid, we construct self-adjoint extensions of the Dirac
Hamiltonian using von Neumann's theory of deficiency indices. We find
self-adjoint extensions of the Dirac Hamiltonian and boundary conditions at
the AB solenoid. Besides, for the first time, solutions of the Dirac
equation in the magnetic-solenoid field with a finite radius solenoid were
found. We study the structure of these solutions and their dependence on the
behavior of the magnetic field inside the solenoid. Then we exploit the
latter solutions to specify boundary conditions for the magnetic-solenoid
field with Aharonov-Bohm solenoid.
\end{abstract}

\section{Introduction}

The present article is a natural continuation of the works \cite
{L83,BGS86,SS89,BGT01} where solutions of the Schr\"{o}dinger, Klein-Gordon,
and Dirac equations in the superposition of the Aharonov-Bohm (AB) field
(the field of an infinitely long and infinitesimally thin solenoid) and a
collinear uniform magnetic field were studied. In what follows, we call the
latter superposition the magnetic-solenoid field. In particular, in the
paper \cite{BGT01} solutions of the Dirac equation in the magnetic-solenoid
field in $2+1$ and $3+1$ dimensions were studied in detail. Then, in \cite
{BGLT01}, these solutions were used to calculate various characteristics of
the particle radiation in such a field. In fact, the AB effect in
synchrotron radiation was investigated. However, a number of important and
interesting aspects related to the rigorous treatment of the solutions of
the Dirac equation in the magnetic-solenoid field were not considered. One
ought to say that in the work \cite{BGT01} it was pointed out that a
critical subspace exists where the Hamiltonian of the problem is not
self-adjoint. But the corresponding self-adjoint extensions of the
Hamiltonian were not studied. The completeness of the solutions was not
considered from this point of view as well.

One has to remark that even for the pure AB field it was not simple to solve
the two aforementioned problems. First, the construction of self-adjoint
extensions of the nonrelativistic Hamiltonian in the AB field was studied in
detail in \cite{T74}. In the work \cite{T74} solutions in the regularized AB
field were thoroughly considered as well.{\LARGE \ }The need to consider
self-adjoint extensions of the Dirac Hamiltonian in the pure AB field in $%
2+1 $ dimensions was recognized in \cite{GJ89,G89}. The interaction between
the magnetic momentum of a charged particle and the AB field essentially
changes the behavior of the wave functions at the magnetic string \cite
{G89,H90,H91}. It was shown that a one-parameter family of boundary
conditions at the origin arises. Self-adjoint extensions of the Dirac
Hamiltonian in $3+1$ dimensions were found in \cite{VGS91}. The works \cite
{AJS95,AJS96} present an alternative method of treating the Hamiltonian
extension problem in $2+1$ and in $3+1$ dimensions. It was shown in \cite
{MT93} that in $2+1$ dimensions only two values of the extension parameter
correspond to the presence of the point-like magnetic field at the origin.
Thus, other values of the parameter correspond to additional contact
interactions. One possible boundary condition was obtained in \cite
{H90,AMW89,FL91} by a specific regularization of the Dirac delta function,
starting from a model in which the continuity of both components of the
Dirac spinor is imposed at a finite radius, and then this radius is shrunk
to zero. Other extensions in $2+1$ and $3+1$ dimensions were constructed in
the works \cite{BFS99,BFKS00,FK01} by imposing spectral boundary conditions
of the Atiyah-Patodi-Singer type \cite{APS75} (MIT boundary conditions) at a
finite radius, and then the zero-radius limit is taken. In the works \cite
{CP94,ACP01} it was shown that, given certain relations between the
extension parameters, it is possible to find the most general domain where
the Hamiltonian and the helicity operator are self-adjoint. The bound state
problem for particles with magnetic moment in the AB potential was
considered in detail in the works \cite{BV93,VB94,BV94}. The physically
motivated boundary conditions for the particle scattering on the AB field
and a Coulomb center was studied in \cite{CP93}.

The study of similar problems in the magnetic-solenoid field is a nontrivial
task. Indeed, the presence of the uniform magnetic field changes the energy
spectrum of the spinning particle from continuous to discrete. Thus, the
boundary conditions that were obtained for a continuous spectrum cannot be
automatically used for the discrete spectrum. By analogy with the pure AB
field it is important to consider the regularized magnetic-solenoid field
(we call the regularized magnetic-solenoid field the superposition of a
uniform magnetic field and the regularized AB field). Here one has to study
solutions of the Dirac equation in such a field. The latter problem was not
solved before, and is of particular interest regardless of the extension
problem in the AB field. One ought to say that the Pauli equation in the
magnetic-solenoid field was recently studied in \cite{T00,C00}.

In the present article we consider the Dirac equation in the general
magnetic-solenoid field (the uniform magnetic field and the AB field may
have both the same and opposite directions) and in the regularized
magnetic-solenoid field. First we construct self-adjoint extensions of the
Dirac Hamiltonian using von Neumann's theory of deficiency indices. We
demonstrate how to reduce the $\left( 3+1\right) $-dimensional problem to
the $\left( 2+1\right) $-dimensional one by a proper choice of the spin
operator. We find self-adjoint extensions of the Dirac Hamiltonian in both
above dimensions and boundary conditions at the AB solenoid. Then, we study
properties of the corresponding solutions and energy spectra. We discuss the
spectrum dependence upon the extension parameter. In the regularized
magnetic-solenoid field, we find for the first time solutions of the Dirac
equation. We study the structure of these solutions and their dependence on
the behavior of the magnetic field inside the solenoid. Then we use these
solutions to specify boundary conditions for the singular magnetic-solenoid
field. To this end, we consider the zero-radius limit of the solenoid. One
ought to say that the problem of the Hamiltonian extension in a particular
case of the magnetic-solenoid field (both fields have the same directions)
was considered in \cite{ESV02} (scalar case) and in \cite{FP01} (spinning
case in $2+1$ dimensions). However, the $3+1$ dimensional spinning problem
was not studied as well as the relation of the extensions with the
regularized problem.

\section{Exact solutions}

Consider the Dirac equation ($c=\hbar =1$) in $\left( 3+1\right) $ and $%
\left( 2+1\right) $ dimensions, 
\begin{equation}
i\partial _{0}\Psi =H\Psi ,\;H=\gamma ^{0}\left( \mbox{\boldmath$\gamma$%
\unboldmath}\mathbf{P}+M\right) \,.  \label{abe1}
\end{equation}
Here $\gamma ^{\nu }=\left( \gamma ^{0},\mbox{\boldmath$\gamma$\unboldmath}%
\right) ,\;\mbox{\boldmath$\gamma$\unboldmath}=\left( \gamma ^{k}\right) $,\ 
$P_{k}=i\partial _{k}-qA_{k},$ $k=1,2,$ for $2+1$, and $k=1,2,3$, for $3+1$,$%
\;\nu =\left( 0,k\right) ;\;q$ is an algebraic charge, for electrons $q=-e<0$%
. As an external electromagnetic field we take the magnetic-solenoid field.
The magnetic-solenoid field is a collinear superposition of a constant
uniform magnetic field $B$ and the Aharonov-Bohm field $B^{AB}$ (the AB
field is a field of an infinitely long and infinitesimally thin solenoid).
The complete Maxwell tensor has the form: 
\begin{equation*}
F_{\lambda \nu }=\overline{B}\left( \delta _{\lambda }^{2}\delta _{\nu
}^{1}-\delta _{\lambda }^{1}\delta _{\nu }^{2}\right) ,\;\overline{B}%
=B^{AB}+B\,.
\end{equation*}
The AB\ field is singular at $r=0$, 
\begin{equation*}
B^{AB}=\Phi \delta (x^{1})\delta (x^{2})\,.
\end{equation*}
The AB\ field creates the magnetic flux $\Phi $. It is convenient to present
this flux as: 
\begin{equation}
\Phi =\left( l_{0}+\mu \right) \Phi _{0},\Phi _{0}=2\pi /e\;,  \label{abex1}
\end{equation}
where $l_{0}$ is integer, and $0\leq \mu <1$.

If we use the cylindric coordinates $\varphi ,r:\;x^{1}=r\cos \varphi $, $%
x^{2}=r\sin \varphi $, then the potentials have the form, 
\begin{eqnarray}
&&A_{0}=0,\;eA_{1}=\left[ l_{0}+\mu +A\left( r\right) \right] \frac{\sin
\varphi }{r},\;eA_{2}=-\left[ l_{0}+\mu +A\left( r\right) \right] \frac{\cos
\varphi }{r}\;,  \notag \\
&&\,(A_{3}=0\;\mathrm{in\;}3+1{}),\;\;A\left( r\right) =eBr^{2}/2\;.
\label{abe2}
\end{eqnarray}

\subsection{Solutions in 2+1 dimensions}

First, we consider the problem in $2+1$ dimensions. In $2+1$ dimensions
there are two non-equivalent representations for $\gamma $-matrices: 
\begin{equation*}
\gamma ^{0}=\sigma ^{3},\;\gamma ^{1}=i\sigma ^{2},\;\gamma ^{2}=-i\sigma
^{1}\zeta ,\;\;\zeta =\pm 1\,,
\end{equation*}
where the ''polarizations'' $\zeta =\pm 1$ correspond to ''spin up'' and
''spin down'', respectively, $\mbox{\boldmath$\sigma$\unboldmath}=\left(
\sigma ^{i}\right) $ are Pauli matrices. In the stationary case, we may
select the following form for the spinors $\Psi (x)$, 
\begin{equation}
\Psi (x)=\exp \left\{ -i\varepsilon x^{0}\right\} \psi _{\varepsilon
}^{(\zeta )}\left( x_{\perp }\right) \,,\;\zeta =\pm 1,\;x_{\perp }=\left(
0,x^{1},x^{2}\right) \,.  \label{abe6}
\end{equation}
Then the Dirac equation in both representations implies: 
\begin{equation}
\left( \mbox{\boldmath$\sigma$\unboldmath}\mathbf{P}_{\perp }+M\sigma
^{3}\right) \psi _{\varepsilon }^{(1)}(x_{\perp })=\varepsilon \psi
_{\varepsilon }^{(1)}(x_{\perp }),\;P_{\perp }=\left( 0,P_{1},P_{2}\right)
\,,  \label{abe7a}
\end{equation}
\begin{equation}
\left( \sigma ^{1}\mbox{\boldmath$\sigma$\unboldmath}\mathbf{P}_{\perp
}\sigma ^{1}+M\sigma ^{3}\right) \psi _{\varepsilon }^{(-1)}(x_{\perp
})=\varepsilon \psi _{\varepsilon }^{(-1)}(x_{\perp })\,.  \label{abe7b}
\end{equation}
We remark that the energy eigenvalues can be positive, $\varepsilon
={}_{+}\varepsilon >0$ , or negative, $\varepsilon ={}_{-}\varepsilon <0$.
One can see that 
\begin{equation}
\psi _{\varepsilon }^{(-1)}(x_{\perp })=\sigma ^{2}\psi _{-\varepsilon
}^{(1)}(x_{\perp })\,.  \label{abe8}
\end{equation}
Further, we are going to use the representation defined by $\zeta =1$.

As the total angular momentum operator we select $J=-i\partial _{\varphi
}+\sigma ^{3}/2$ that is the dimensional reduction of the operator $J^{3}$
in $3+1$ dimensions. The operator $J$ commutes with the Hamiltonian $H$.
Then the spinors $\psi _{\varepsilon }^{(1)}$ have to satisfy Eq. (\ref
{abe7a}) and the equation 
\begin{equation}
J\psi _{\varepsilon }^{(1)}\left( x_{\perp }\right) =\left( l-l_{0}-\frac{1}{%
2}\right) \psi _{\varepsilon }^{(1)}\left( x_{\perp }\right) \,,\;l\in 
\mathbb{Z}\,.  \label{abe10}
\end{equation}
Presenting the spinors $\psi _{\varepsilon }^{(1)}$ in the form 
\begin{equation}
\psi _{\varepsilon }^{(1)}\left( x_{\perp }\right) =g_{l}(\varphi )\psi
_{l}\left( r\right) ,\;g_{l}(\varphi )=\frac{1}{\sqrt{2\pi }}\exp \left\{
i\varphi \left( l-l_{0}-\frac{1}{2}\left( 1+\sigma ^{3}\right) \right)
\right\} \,,  \label{abe11}
\end{equation}
we find that the radial spinor $\psi _{l}\left( r\right) $ obeys the
equation 
\begin{eqnarray}
&&h\psi _{l}(r)=\varepsilon \psi _{l}(r),\;\;h=\Pi +\sigma ^{3}M\,,
\label{abe12} \\
&&\Pi =-i\left\{ \partial _{r}+\frac{\sigma ^{3}}{r}\left[ \mu +l-\frac{1}{2}%
\left( 1-\sigma ^{3}\right) +A\left( r\right) \right] \right\} \sigma ^{1}\;.
\label{abe13}
\end{eqnarray}
Here $h$ is the radial Hamiltonian, $\Pi $ defines the action of the spin
projection operator on the radial spinor in the subspace with a given $l$, 
\begin{equation*}
\mbox{\boldmath$\sigma$\unboldmath}\mathbf{P}_{\perp }g_{l}(\varphi )\psi
_{l}\left( r\right) =g_{l}(\varphi )\Pi \psi _{l}\left( r\right) \,.
\end{equation*}

It is convenient to present the radial spinor in the following form 
\begin{equation}
\psi _{l}(r)=\left[ \sigma ^{3}\left( \varepsilon -\Pi \right) +M\right]
u_{l}(r)\,,  \label{abe15}
\end{equation}
where 
\begin{eqnarray}
&&u_{l}(r)=\sum_{\sigma =\pm 1}c_{\sigma }u_{l,\sigma }(r)\,,\;u_{l,\sigma
}(r)=\phi _{l,\sigma }(r)\upsilon _{\sigma }\,,  \notag \\
&&\upsilon _{1}=\left( 
\begin{array}{l}
1 \\ 
0
\end{array}
\right) ,\;\;\upsilon _{-1}=\left( 
\begin{array}{l}
0 \\ 
1
\end{array}
\right) ,  \label{abe16}
\end{eqnarray}
and $c_{\sigma }$ are some constants. It follows from (\ref{abe12}) that $%
\Pi ^{2}u=\left( \varepsilon ^{2}-M^{2}\right) u$, therefore the radial
functions $\phi _{l,\sigma }(r)$ satisfy the following equation: 
\begin{eqnarray}
&&\left\{ \rho \frac{d^{2}}{d\rho ^{2}}+\frac{d}{d\rho }-\frac{\rho }{4}+%
\frac{1}{2}\left[ \frac{\omega }{\gamma }-\xi \left( \mu +l-\frac{1}{2}%
\left( 1-\sigma \right) \right) \right] -\frac{\nu ^{2}}{4\rho }\right\}
\phi _{l,\sigma }(r)=0\,,  \label{abe17} \\
&&\rho =\gamma r^{2}/2,\;\;\gamma =e\left| B\right| ,\;\;\xi =\mathrm{sgn}%
B,\;\;\nu =\mu +l-\left( 1+\sigma \right) /2,\;\omega =\varepsilon
^{2}-M^{2}\,.  \notag
\end{eqnarray}
Solutions of the equation (\ref{abe17}) were studied in \cite{BGT01}. Taking
into account these results, we get:

For any $l$, there exist a set of regular\footnote{%
Here we use the terms ''regular'', ''irregular'' at $r=0$ in the following
sense. We call a function to be regular if it behaves as $r^{c}$ at $r=0$
with $c\geq 0$, and irregular if $c<0$. We call a spinor to be regular when
all its components are regular, and irregular when at least one of its
components is irregular.} at $r=0$ solutions $\phi _{l,\sigma }=(\phi
_{m,l,\sigma }\,$, $m=0,1,2,\ldots )$, 
\begin{equation}
\phi _{m,l,\sigma }(r)=I_{m+|\nu |,m}\left( \rho \right) \,.  \label{abe19}
\end{equation}
Here $I_{n,m}(\rho )$ are the Laguerre functions that are presented in
Appendix A.

For $l=0$ there exist solutions irregular at $r=0$. A general irregular
solution for $l=0$, $\mu \neq 0$ reads: 
\begin{eqnarray}
&&\phi _{\omega ,\sigma }(r)=\psi _{\lambda ,\alpha }(\rho )=\rho
^{-1/2}W_{\lambda ,\alpha /2}(\rho )\,,  \notag \\
&&\alpha =\mu -\left( 1+\sigma \right) /2,\;2\lambda =\omega /\gamma -\xi 
\left[ \mu -\left( 1-\sigma \right) /2\right] \,,  \label{abe25}
\end{eqnarray}
where $W_{\lambda ,\alpha /2}$ are the Whittaker functions (see \cite{GR94}
, 9.220.4). The spinors in (\ref{abe12}) constructed with the help of the
latter functions\ are square integrable for arbitrary complex $\lambda $.
The functions $\psi _{\lambda ,\alpha }$ were studied in detail in \cite
{BGT01}, some important relations for these functions are presented in
Appendix A. We see that interpretation of $\omega $\ as energy is impossible
for complex $\lambda $. For real $\lambda $\ there exist a set of solutions (%
\ref{abe25}) which can be expressed in terms of the Laguerre functions with
integer indices: 
\begin{eqnarray}
\phi _{m,+1}^{ir}(r) &=&I_{m+\mu -1,m}\left( \rho \right) \,,\;\sigma
=+1{},\;m=0,1,2,\ldots ,  \notag \\
\phi _{m,-1}^{ir}(r) &=&I_{m-\mu ,m}\left( \rho \right) \,,\;\sigma
=-1,\;m=0,1,2,\ldots \,.  \label{abe20}
\end{eqnarray}
All the corresponding solutions $\psi _{l}(r)$ of Eq. (\ref{abe12}) are
square integrable on the half-line with the measure $rdr$. The Laguerre
functions in Eqs. (\ref{abe19}), (\ref{abe20}) are expressed via the
Laguerre polynomials.

Eigenvalues $\omega $ and the form of spinors depend on $\mathrm{sgn}B$.
Below we present the results for $B>0$. The results for $B<0$ can not be
obtained trivially from ones for $B>0$. We present them in Appendix B. The
spectrum of $\omega $ corresponding to the functions $\phi _{m,l,\sigma }(r)$
reads 
\begin{equation}
\omega =\left\{ 
\begin{array}{l}
2\gamma \left( m+l+\mu \right) ,\;{}l-\left( 1+\sigma \right) /2\geq 0 \\ 
2\gamma \left( m+\left( 1+\sigma \right) /2\right) ,\;{}l-\left( 1+\sigma
\right) /2<0
\end{array}
\right. ,  \label{abe21}
\end{equation}
and the spectrum of $\omega $ corresponding to the functions $\phi
_{m,\sigma }^{ir}(r)$ reads 
\begin{equation}
\omega =\left\{ 
\begin{array}{l}
2\gamma \left( m+\mu \right) ,\;\sigma =1 \\ 
2\gamma m,\;\sigma =-1
\end{array}
\right. .  \label{abe22}
\end{equation}

We demand the spinors $u_{l}\left( r\right) $ to be eigenvector for $\Pi $,
such that the functions $u_{m,l,\pm }$ have to obey the equation 
\begin{equation}
\Pi u_{m,l,\pm }(r)=\pm \sqrt{\omega }u_{m,l,\pm }(r)\,.  \label{abe26}
\end{equation}
Then we can specify the coefficients in (\ref{abe16}).

In the case $\omega =0$, 
\begin{equation}
u_{0,l}(r)=\left( 
\begin{array}{l}
0 \\ 
\phi _{0,l,-1}(r)
\end{array}
\right) ,\;{}l\leq -1;\;\;\;u_{0}^{I}(r)=\left( 
\begin{array}{l}
0 \\ 
\phi _{0,-1}^{ir}(r)
\end{array}
\right) ,\;{}l=0\;.  \label{abe27}
\end{equation}
That can be easily seen from the relations (\ref{4.11}) - (\ref{4.16}) for
the Laguerre functions $I_{n,m}(\rho )$ .

In the case $\omega \neq 0$, 
\begin{eqnarray}
u_{m,l,\pm }(r) &=&\left( 
\begin{array}{l}
\phi _{m,l,+1}(r) \\ 
\pm i\phi _{m,l,-1}(r)
\end{array}
\right) \,,\mathrm{\;}l\geq 1,\;\omega =2\gamma \left( m+l+\mu \right) \,, 
\notag \\
u_{m+1,l,\pm }(r) &=&\left( 
\begin{array}{l}
\phi _{m,l,+1}(r) \\ 
\mp i\phi _{m+1,l,-1}(r)
\end{array}
\right) \,,\mathrm{\;}l\leq -1,\;\omega =2\gamma \left( m+1\right) \,, 
\notag \\
u_{m+1,\pm }^{I}(r) &=&\left( 
\begin{array}{l}
\phi _{m,0,+1}(r) \\ 
\mp i\phi _{m+1,-1}^{ir}(r)
\end{array}
\right) ,\mathrm{\;}l=0,\;\omega =2\gamma \left( m+1\right) \,,  \notag \\
u_{m,\pm }^{II}(r) &=&\left( 
\begin{array}{l}
\phi _{m,+1}^{ir}(r) \\ 
\pm i\phi _{m,0,-1}(r)
\end{array}
\right) ,\;l=0,\;\omega =2\gamma \left( m+\mu \right) \,.  \label{abe28}
\end{eqnarray}

For $\omega \neq 0$, we construct solutions of the Dirac equation using the
spinors $u$ corresponding to the positive eigenvalues of the operator $\Pi $%
. These solutions have the form, 
\begin{eqnarray}
&&\psi _{m,l}(r)=N\left[ \sigma ^{3}\left( \varepsilon -\sqrt{\omega }%
\right) +M\right] u_{m,l,+}(r),\;l\neq 0\,,  \notag \\
&&\psi _{m}^{I,II}(r)=N\left[ \sigma ^{3}\left( \varepsilon -\sqrt{\omega }%
\right) +M\right] u_{m,+}^{I,II}(r),\;l=0\,,  \label{abe31}
\end{eqnarray}
where $N$ is a normalization constant. Substituting (\ref{abe31}) into (\ref
{abe12}) we obtain two types of states corresponding to particles $%
\,_{+}\psi $ and antiparticles $\,_{-}\psi $ with $\varepsilon ={}_{\pm
}\varepsilon =\pm \sqrt{M^{2}+\omega }$, respectively. The particle and
antiparticle spectra are symmetric, that is $|_{+}\varepsilon
|=|_{-}\varepsilon |$, for given quantum numbers $m$, $l$.

Consider the case $\omega =0$. As it follows from (\ref{abe12}), (\ref{abe27}%
) only negative energy solutions (antiparticles) are possible. They coincide
with the corresponding spinors $u$ up to a normalization constant 
\begin{equation}
_{-}\psi _{0,l}(r)=Nu_{0,l}(r){},\;l\leq -1;\;\;\;_{-}\psi
_{0}^{I}(r)=Nu_{0}^{I}(r){},\;l=0\;.  \label{abe32}
\end{equation}
Thus, only antiparticles have the rest energy level. The particle\ lowest
energy level for $l\leq 0$ is $_{+}\varepsilon =\sqrt{M^{2}+2\gamma }$.

All the radial spinors $\psi _{m,l}(r)$ are orthogonal for different $m$.
The same is true both for the spinors $\psi _{m}^{I}$ and $\psi _{m}^{II}$.
In the general case, the spinors of the different types are not orthogonal.
By the help of Eq. (\ref{4.41}) of Appendix A, one can prove this fact and
at the same time calculate the normalization factor which has the same form
for all types of the spinors, 
\begin{equation}
N=\frac{\sqrt{\gamma }}{\sqrt{2\left[ \left( \varepsilon -\sqrt{\omega }%
\right) ^{2}+M^{2}\right] }}\,.  \label{abe33b}
\end{equation}

Besides, on the subspace $l=0$ there are solutions of Eq. (\ref{abe12}) that
are expressed via the functions $\psi _{\lambda ,\alpha }(\rho )$ (\ref
{abe25}). We present these solutions as follows, 
\begin{eqnarray}
\psi _{\omega }(r) &=&\left[ \sigma ^{3}\left( \varepsilon -\Pi \right) +M%
\right] u_{\omega }(r)\,,  \notag \\
u_{\omega }(r) &=&c_{1}u_{\omega ,+1}(r)+c_{-1}u_{\omega ,-1}(r),\;u_{\omega
,\sigma }(r)=\phi _{\omega ,\sigma }(r)v_{\sigma }\,.  \label{abe34}
\end{eqnarray}
Using the relations (\ref{4.145}) for the functions $\psi _{\lambda ,\alpha
}(\rho )$, we obtain the useful expressions 
\begin{equation}
\Pi u_{\omega ,1}(r)=i\sqrt{2\gamma }u_{\omega ,-1}(r),\;\Pi u_{\omega
,-1}(r)=-i\frac{\omega }{\sqrt{2\gamma }}u_{\omega ,+1}(r)\,,  \label{abe35}
\end{equation}
By the help of Eq. (\ref{4.148}) from Appendix A, one can see that the
spinors $\psi _{\omega }(r)$, $\psi _{\omega ^{\prime }}(r)$, $\omega \neq
\omega ^{\prime }$ are not orthogonal in the general case.

The fact that on the subspace $l=0$ (in what follows we call this subspace
the critical subspace, and the subspace $l\neq 0$ the noncritical subspace)
there exist solutions with complex eigenvalues indicates that the radial
Hamiltonian is not self-adjoint, at least on this subspace.

\subsection{Solutions in 3+1 dimensions}

To exploit the symmetry of the problem under $z$ translations, we use the
following representations for $\gamma $-matrices (see \cite{AMW89}), 
\begin{equation*}
\gamma ^{0}=\left( 
\begin{array}{cc}
\sigma ^{3} & 0 \\ 
0 & -\sigma ^{3}
\end{array}
\right) ,\;\gamma ^{1}=\left( 
\begin{array}{cc}
i\sigma ^{2} & 0 \\ 
0 & -i\sigma ^{2}
\end{array}
\right) ,\;\gamma ^{2}=\left( 
\begin{array}{cc}
-i\sigma ^{1} & 0 \\ 
0 & i\sigma ^{1}
\end{array}
\right) ,\;\gamma ^{3}=\left( 
\begin{array}{cc}
0 & I \\ 
-I & 0
\end{array}
\right) \,.
\end{equation*}
In $3+1$ dimensions a complete set of commuting operators can be chosen as
follows ($\gamma ^{5}=-i\gamma ^{0}\gamma ^{1}\gamma ^{2}\gamma ^{3}$), 
\begin{equation}
H,\;P^{3}=-i\partial _{3},\;J^{3}=-i\partial _{\varphi }+\Sigma
^{3}/2,\;S^{3}=\gamma ^{5}\gamma ^{3}\left( M+\gamma ^{3}P^{3}\right) /M\,.
\label{abe3}
\end{equation}
Then we demand the wave function to be eigenvector for these operators, 
\begin{eqnarray}
H\Psi &=&\varepsilon \Psi ,  \label{abe4.1} \\
P^{3}\Psi &=&p^{3}\Psi ,  \label{abe4.2} \\
J^{3}\Psi &=&j^{3}\Psi ,  \label{abe4.3} \\
S^{3}\Psi &=&s\widetilde{M}/M\Psi \,.  \label{abe4.4}
\end{eqnarray}
Here $\widetilde{M}=\sqrt{M^{2}+(p_{3})^{2}}$, $p^{3}$ is $z$-component of
the momentum and $j^{3}$ is $z$-component of the total angular momentum.
Remark that the energy eigenvalues can be positive, $\varepsilon
={}_{+}\varepsilon >0$ , or negative, $\varepsilon ={}_{-}\varepsilon <0$.$%
\; $The eigenvalues $j^{3}$ are half-integer, it is convenient to use the
representation: $j^{3}=\left( l-l_{0}-\frac{1}{2}\right) $, where $l=0,\pm
1,\pm 2,...$ . To specify the spin degree of freedom we select the operator $%
S^{3}$ which is the $z$-component of the polarization pseudovector \cite
{ST68}, 
\begin{equation}
S^{0}=-\frac{1}{2M}\left( H\gamma ^{5}+\gamma ^{5}H\right) ,\;S^{i}=\frac{1}{%
2M}\left( H\Sigma ^{i}+\Sigma ^{i}H\right) \,,  \label{x.3}
\end{equation}
eigenvalues of the\ corresponding spin projections are $s\widetilde{M}/M$, $%
s=\pm 1$.

Then in $3+1$ dimensions one can separate the spin and coordinate variables
and get the following representation for the spinors $\Psi $, 
\begin{eqnarray}
\Psi (x) &=&\exp \left\{ -i\varepsilon x^{0}+ip^{3}x^{3}\right\} \Psi
_{s}(x_{\perp })\,,  \notag \\
\Psi _{s}(x_{\perp }) &=&N\left( 
\begin{array}{c}
\left[ 1+\left( p^{3}+s\widetilde{M}\right) /M\right] \psi _{\varepsilon
,s}(x_{\perp }) \\ 
\left[ -1+\left( p^{3}+s\widetilde{M}\right) /M\right] \psi _{\varepsilon
,s}(x_{\perp })
\end{array}
\right) \,.  \label{abe4}
\end{eqnarray}
Here $\psi _{\varepsilon ,s}(x_{\perp })$ are two-component spinors, $%
x_{\perp }=\left( 0,x^{1},x^{2},0\right) $, $N$ is a normalization factor.

As a result, the equation (\ref{abe4.1}) is reduced to the equation 
\begin{equation}
\left( \mathbf{\mbox{\boldmath$\sigma$\unboldmath}P}_{\perp }+s\widetilde{M}%
\sigma ^{3}\right) \psi _{\varepsilon ,s}(x_{\perp })=\varepsilon \psi
_{\varepsilon ,s}(x_{\perp }),\;P_{\perp }=\left( 0,P_{1},P_{2},0\right) \,.
\label{abe5}
\end{equation}
Presenting $\psi _{\varepsilon ,s}\left( x_{\perp }\right) $ in the form 
\begin{equation}
\psi _{\varepsilon ,s}\left( x_{\perp }\right) =g_{l}(\varphi )\psi
_{l,s}\left( r\right) \,,  \label{abe5n0}
\end{equation}
where $g_{l}\left( \varphi \right) $\ is given by Eq. (\ref{abe11}), one
comes to the radial equation 
\begin{equation}
h_{s}\psi _{l,s}(r)=\varepsilon \psi _{l,s}(r),\;h_{s}=\Pi +s\widetilde{M}%
\sigma ^{3}\,,  \label{abe5n1}
\end{equation}
where $h_{s}$\ is the radial Hamiltonian acting on the subspace with the
spin quantum number $s$, $\Pi $\ is given by Eq. (\ref{abe13}). We remark
that 
\begin{equation}
\psi _{\varepsilon ,-1}(x_{\perp })=\sigma ^{3}\psi _{-\varepsilon
,1}(x_{\perp })\,.  \label{abe5b}
\end{equation}
One can see that at fixed $s$ and $p^{3}$, Eq. (\ref{abe5}) is similar to
Eq. (\ref{abe7a}) in $2+1$ dimensions. Thus, after separation the angular
variable with the help of (\ref{abe11}), the radial spinor\ (\ref{abe5n0}), $%
\psi _{l,+1}\left( r\right) $, can be obtained from the radial spinor (\ref
{abe11}), $\psi _{l}\left( r\right) $, with the substitution $M$ by $%
\widetilde{M}$. The same is true for the particular case $l=0$. Here the
radial spinor\ $\psi _{\omega ,+1}\left( r\right) $, can be obtained from
the radial spinor (\ref{abe34}), $\psi _{\omega }\left( r\right) $.

Using the results for $\left( 2+1\right) $-dimensional case, one concludes
that in the critical subspace complex eigenvalues of Eq. (\ref{abe4.1})
exist. That means the Hamiltonian in $3+1$ dimensions is not self-adjoint.

\section{Self-adjoint extensions}

As well-known \cite{GJ89,G89,AMW89}, the radial Hamiltonian in the pure AB
field requires a self-adjoint extension for the critical subspace $l=0$. As
a result \cite{G89} one gets a one-parameter family of acceptable boundary
conditions. In the case of our interest, the external background is more
complicated, it includes besides the AB field a uniform magnetic field. The
wave functions and the spectrum in such a background differ in a nontrivial
manner from ones in the pure AB field. Thus, the problem of self-adjoint
extension of the Dirac Hamiltonian in such a background, which is considered
below, is not trivial.

\subsection{Extensions in 2+1 dimensions}

First, we study the $\left( 2+1\right) $-dimensional case. To this end we
use the standard theory of von Neumann deficiency indices \cite{RS72}. The $%
\left( 2+1\right) $-dimensional case was formally considered in \cite{FP01}.
We reproduce calculations for the $\left( 2+1\right) $-dimensional case in
terms of the functions of Sect. 2. We generalize the results of \cite{FP01}
for arbitrary sign of $B$ that allows to determine the non-trivial spectrum
dependence on the signs of $B$, $\Phi $. The $\left( 2+1\right) $%
-dimensional case results are necessary in order to extend this result to
the $\left( 3+1\right) $-dimensional case.

The Hamiltonian (\ref{abe1}) acts on the space of two-spinors $\psi \left(
x_{\bot }\right) $. The angular variable can be separated, Eq. (\ref{abe12}%
), that enables us to single out the radial Hamiltonian $h$ (\ref{abe12})
which acts on the space of two-spinors $\psi \left( r\right) $. We start
with the choice of the domain of definition of $h$, $\mathcal{D}\left(
h\right) $. Let $\mathcal{D}\left( h\right) $ be the space of absolutely
continuous square integrable on the half-line (with the measure $rdr)$ and
regular at the origin functions. One can make sure that $h$ is symmetric on
the domain $\mathcal{D}\left( h\right) $. To determine whether the
Hamiltonian is self-adjoint we have to define its deficiency indices, $%
n_{\pm }\left( h\right) =\dim \left( \mathcal{D}^{\pm }\right) $, $\mathcal{D%
}^{\pm }=$\textrm{Ker}$\left( h^{\dagger }\mp iM\right) $, where $h^{\dagger
}$ is the adjoint of $h$. That is we have to find the number of linearly
independent solutions of the equations, 
\begin{eqnarray}
&&h^{\dagger }\psi ^{\pm }(r)=\pm iM\psi ^{\pm }(r),\;h^{\dagger }=\Pi
^{\dagger }+\sigma ^{3}M\,,  \label{abe39} \\
&&\Pi ^{\dagger }=-i\left\{ \partial _{r}+\frac{\sigma ^{3}}{r}\left[ \mu +l-%
\frac{1}{2}\left( 1-\sigma ^{3}\right) +A\left( r\right) \right] \right\}
\sigma ^{1}\,.  \label{abe40}
\end{eqnarray}
Here $M$ is introduced by dimensional reasons. For both cases ($\psi ^{\pm
}\left( r\right) $ ) there exist only one linear independent
square-integrable solution, for $l=0$, that reads, 
\begin{eqnarray}
\psi ^{\pm }(r) &=&N\left( 
\begin{array}{l}
\phi _{1}(r) \\ 
\pm e^{\pm i\pi /4}\frac{\sqrt{\gamma }}{M}\phi _{-1}(r)
\end{array}
\right) ,\;B>0\,,  \label{abe41} \\
\psi ^{\pm }(r) &=&N\left( 
\begin{array}{l}
\phi _{1}(r) \\ 
\pm e^{\pm i\pi /4}\frac{M}{\sqrt{\gamma }}\phi _{-1}(r)
\end{array}
\right) ,\;B<0\,,  \label{abe42}
\end{eqnarray}
where, using (\ref{abe25}), 
\begin{equation*}
\phi _{\sigma }(r)=\psi _{\lambda ,\alpha }(\rho ),\;2\lambda
=-2M^{2}/\gamma -\xi \left( \mu -\left( 1-\sigma \right) /2\right) ,\;\sigma
=\pm 1\,.
\end{equation*}
Thus, on the non-critical subspace the deficiency indices are $(0,0)$, and
on the critical subspace the deficiency indices are $(1,1)$. Therefore, on
the non-critical subspace the radial Hamiltonian is (essentially)
self-adjoint, and on the critical subspace the radial Hamiltonian has
self-adjoint extensions. Besides, there exist the isometry from $\mathcal{D}%
^{+}$ into $\mathcal{D}^{-}$, $\psi ^{+}(r)\rightarrow e^{i\Omega }\psi
^{-}(r),\;\Omega \in \mathbb{R}$. According to von Neumann's theory, the
extensions of a closed symmetric operator\footnote{%
We remark, that every symmetric operator has a closure, and the operator and
its closure have the same closed extensions \cite{RS72}.} are in one-to-one
correspondence with a set of isometries. Thus, self-adjoint extensions of
the original operator $h$ form the one-parameter family labelled by the
parameter $\Omega $, $h^{\Omega }$. The domain of $h^{\Omega }$ reads, 
\begin{equation}
\mathcal{D}\left( h^{\Omega }\right) =\left\{ \chi \left( r\right) =\psi
\left( r\right) +c\left( \psi ^{+}\left( r\right) +e^{i\Omega }\psi
^{-}\left( r\right) \right) :\;\psi \left( r\right) \in \mathcal{D}\left(
h\right) \right\} ,\;c\in \mathbb{C}\mathbf{\,},  \label{abe43n}
\end{equation}
where $\chi $ is a two component spinor, $\chi =\left( \chi ^{1},\chi
^{2}\right) $. The behavior of the functions from $\mathcal{D}\left(
h^{\Omega }\right) $ at $r\rightarrow 0$ is defined by the behavior of $\chi
\left( r\right) $. Using the behavior (\ref{4.147}) of the function $\psi
_{\lambda ,\alpha }(\rho )$ at small $\rho $, we find, 
\begin{equation}
\lim_{r\rightarrow 0}\frac{\chi ^{1}\left( r\right) \left( Mr\right) ^{1-\mu
}}{\chi ^{2}\left( r\right) \left( Mr\right) ^{\mu }}=\left\{ 
\begin{array}{c}
\frac{i2^{1-\mu }\Gamma (1-\mu )\Gamma (\mu +M^{2}/\gamma )}{\left( \tan 
\frac{\Omega }{2}-1\right) \Gamma (\mu )\Gamma (1+M^{2}/\gamma )}\left( 
\frac{M^{2}}{\gamma }\right) ^{1-\mu },\;B>0\,, \\ 
\frac{i2^{1-\mu }\Gamma (1-\mu )\Gamma (1+M^{2}/\gamma )}{\left( \tan \frac{%
\Omega }{2}-1\right) \Gamma (\mu )\Gamma (1-\mu +M^{2}/\gamma )}\left( \frac{%
M^{2}}{\gamma }\right) ^{-\mu }\,,\;B<0
\end{array}
\right. \,.  \label{abe43}
\end{equation}
One can verify that the limit $\gamma \rightarrow 0$ of the right hand sides
of (\ref{abe43}) coincides with the corresponding expression obtained in 
\cite{G89} in the case of pure AB field. For our purposes it is convenient
to pass from the parametrization by $\Omega $ to the parametrization by the
angle $\Theta $, $0\leq \Theta <2\pi $,\ such that 
\begin{equation}
\lim_{r\rightarrow 0}\frac{\chi ^{1}\left( r\right) \left( Mr\right) ^{1-\mu
}}{\chi ^{2}\left( r\right) \left( Mr\right) ^{\mu }}=i\tan \left( \frac{\pi 
}{4}+\frac{\Theta }{2}\right) \,.  \label{abe44}
\end{equation}
To guarantee the self-adjointness of the Hamiltonian one has to demand the
functions of its domain to satisfy Eq. (\ref{abe44}).

Thus, the solutions (\ref{abe34}) obtained in Sect. 2 must be subjected to
the condition (\ref{abe44}) at $r\rightarrow 0$. Then, with the help of (\ref
{abe35}), (\ref{abe36}) and (\ref{4.147}), we find 
\begin{equation}
\tan \left( \frac{\pi }{4}+\frac{\Theta }{2}\right) =\left\{ 
\begin{array}{c}
-\frac{\left( \varepsilon +M\right) }{M}\,\frac{\Gamma (1-\mu )\Gamma (\mu
-\omega /2\gamma )}{2^{\mu }\Gamma (\mu )\Gamma (1-\omega /2\gamma )}\left( 
\frac{M^{2}}{\gamma }\right) ^{1-\mu },\;B>0\, \\ 
\frac{M}{\left( \varepsilon -M\right) }\,\frac{\Gamma (1-\mu )\Gamma
(1-\omega /2\gamma )}{2^{\mu -1}\Gamma (\mu )\Gamma (1-\mu -\omega /2\gamma )%
}\left( \frac{M^{2}}{\gamma }\right) ^{-\mu },\;B<0
\end{array}
\right. \,.  \label{abe48}
\end{equation}

\subsection{Extensions in 3+1 dimensions}

Now we pass to the $\left( 3+1\right) $-dimensional case. Usually, the
helicity operator $S_{h}=\mathbf{\Sigma P/|P}|$ is used as the spin
operator. It relates to $t$-component of polarization pseudovector (\ref{x.3}%
) as, $S_{h}=S^{0}M/\left| \mathbf{P}\right| $. In this case it is necessary
to find a common domain for two operators: $H$ and $S_{h}$. That is not a
trivial problem even in the special case $p^{3}=0$ \cite{CP94,ACP01}.
Moreover, not for all the extension parameter values of the Hamiltonian
there exists a self-adjoint extension of the operator $S_{h}$. This is the
principal reason of our choice $S^{3}$ as the spin operator.

In $\left( 3+1\right) $-dimensional case the Hamiltonian (\ref{abe1}) acts
on the space of 4-spinors of the form (\ref{abe4}). The Hilbert space of
4-spinors (\ref{abe4}) can be presented as the direct sum of two orthogonal
subspaces with respect to the value of the spin quantum number $s$: $%
\mathcal{D}\left( H\right) =\left\{ \Psi _{+1}\right\} \oplus \left\{ \Psi
_{-1}\right\} $. Then we consider the Hamiltonian (\ref{abe1}) on each of
the subspaces. Using Eqs. (\ref{abe5}), (\ref{abe11}) allows to single out
the radial Hamiltonian $h_{s}$ acting on the subspace with the spin quantum
number $s$.

We choose the domain of definition of $h_{s}$, $\mathcal{D}\left(
h_{s}\right) $ as follows. Let $\mathcal{D}\left( h_{s}\right) $ be the
space of absolutely continuous square integrable on the half-line (with the
measure $rdr)$ and regular at the origin functions. The radial Hamiltonian $%
h_{s}$ is symmetric on the domain $\mathcal{D}\left( h_{s}\right) $. Now we
apply von Neumann's theory of deficiency indices to each of the subspaces.

To define the deficiency indices of operators $h_{s}$ we have to solve the
problem, 
\begin{equation}
h_{s}^{\dagger }\psi _{s}^{\pm }\left( r\right) =\pm is\widetilde{M}\psi
_{s}^{\pm }\left( r\right) ,\;h_{s}^{\dagger }=\Pi ^{\dagger }+s\widetilde{M}%
\sigma ^{3},\;s=\pm 1\,,  \label{abe50x1}
\end{equation}
where $h_{s}^{\dagger }$\ is the adjoint of $h_{s}$, $\Pi ^{\dagger }$ is
given by (\ref{abe40}). One can see that Eq. (\ref{abe50x1}) is similar to
Eq. (\ref{abe39}). Then, using Eqs. (\ref{abe41}) [or (\ref{abe42})], (\ref
{abe5b}) one obtains that for $l=0$ the solutions read, 
\begin{eqnarray}
&&\psi _{s}^{\pm }(r)=N\left( 
\begin{array}{l}
\phi _{s,+1}(r) \\ 
\pm se^{\pm i\pi /4}\frac{\sqrt{\gamma }}{\widetilde{M}}\phi _{s,-1}(r)
\end{array}
\right) ,\;B>0\,,  \label{abe50x3} \\
&&\psi _{s}^{\pm }(r)=N\left( 
\begin{array}{l}
\phi _{s,+1}(r) \\ 
\pm se^{\pm i\pi /4}\frac{\widetilde{M}}{\sqrt{\gamma }}\phi _{s,-1}(r)
\end{array}
\right) ,\;B<0\,,  \label{abe50x4} \\
&&\phi _{s,\sigma }(r)=\psi _{\lambda ,\alpha }(\rho )\,,\;\alpha =\mu
-\left( 1+\sigma \right) /2\,,  \notag \\
&&2\lambda =-2\widetilde{M}^{2}/\gamma -\xi \left( \mu -\left( 1-\sigma
\right) /2\right) \,,\;\sigma =\pm 1\,.  \notag
\end{eqnarray}
whereas for $l\neq 0$ there exist no square integrable solutions. Therefore,
for each subspace $s=\pm 1$ on the non-critical subspace the deficiency
indices are $(0,0)$, and on the critical subspace the deficiency indices are 
$(1,1)$. Thus, on the non-critical subspace the radial Hamiltonian $h_{s}$
is (essentially) self-adjoint, and on the critical subspace it has
self-adjoint extensions.

Using the results of $\left( 2+1\right) $-dimensional case we conclude that
on each subspace $s=\pm 1$ self-adjoint extensions of the radial Hamiltonian 
$h_{s}$ form the one-parameter family labelled by the parameter $\Omega _{s}$%
, $h_{s}^{\Omega _{s}}$. The domain of $h_{s}^{\Omega _{s}}$ on each
subspace reads: 
\begin{equation}
\mathcal{D}\left( h_{s}^{\Omega _{s}}\right) =\left\{ \chi _{s}\left(
r\right) =\psi _{s}\left( r\right) +c\left( \psi _{s}^{+}\left( r\right)
+e^{i\Omega _{s}}\psi _{s}^{-}\left( r\right) \right) :\;\psi _{s}\left(
r\right) \in \mathcal{D}\left( h_{s}\right) \right\} ,\;c\in \mathbb{C}\,,
\label{r12}
\end{equation}
Using the parametrization by the angle $\Theta _{s}$ similar to (\ref{abe44}%
) we define the condition for the functions from domain $\mathcal{D}\left(
h_{s}^{\Omega _{s}}\right) $ at $r\rightarrow 0$\ as follows 
\begin{equation}
\lim_{r\rightarrow 0}\frac{\chi _{s}^{1}\left( r\right) \left( \widetilde{M}%
r\right) ^{1-\mu }}{\chi _{s}^{2}\left( r\right) \left( \widetilde{M}%
r\right) ^{\mu }}=is\tan \left( \frac{\pi }{4}+\frac{\Theta _{s}}{2}\right)
,\;s=\pm 1\,.  \label{r14}
\end{equation}
Therefore, in each subspace$\;s=\pm 1$ solutions $\psi _{\omega ,s}\left(
r\right) $ on the critical subspace must be subjected to the condition (\ref
{r14}) at $r\rightarrow 0$. Thus, in $3+1$ dimensions there exist the
two-parameter family of self-adjoint extensions of the Hamiltonian.

\subsection{Spectra of self-adjoint extensions}

Let us study spectra of the self-adjoint extensions $h^{\Omega }$. To this
end we have to solve the transcendental equations (\ref{abe48}) for $\omega $
considering two branches of $\varepsilon $, one for particles and another
one for antiparticles, $_{\pm }\varepsilon =\pm \sqrt{M^{2}+\omega }$.
Introducing the notations 
\begin{eqnarray}
\,\omega &=&2\gamma x,\;\;x=\,_{\varsigma }x=\left( \,_{\varsigma
}\varepsilon ^{2}-M^{2}\right) /2\gamma \,,\;Q\left( x\right) =\frac{%
\varepsilon }{M}+1\,,\;\varsigma =\pm \,,  \notag \\
\,\,\eta &=&\frac{2^{\mu }\Gamma (\mu )}{\Gamma (1-\mu )}\tilde{\eta}\left(
\mu \right) ,\;\;\tilde{\eta}\left( \mu \right) =-\tan \left( \frac{\pi }{4}+%
\frac{\Theta }{2}\right) \left( \frac{\gamma }{M^{2}}\right) ^{1-\mu }\,,
\label{abe50n}
\end{eqnarray}
we can rewrite Eq. (\ref{abe48}) for $B>0$ as follows, 
\begin{equation}
Q\left( \,_{\varsigma }x\right) \frac{\Gamma (\mu -\,_{\varsigma }x)}{\Gamma
(1-\,_{\varsigma }x)}=\eta \,.  \label{abe501}
\end{equation}
Having $\omega $ for $B>0,$ one can obtain $\omega $ for $B<0$ making the
transformation 
\begin{equation*}
\varsigma \rightarrow -\varsigma ,{\LARGE \ \;}\tilde{\eta}\left( \mu
\right) \rightarrow 1/\tilde{\eta}\left( \mu \right) ,\;\;\mu \rightarrow
1-\mu \,.
\end{equation*}
Therefore, below we consider the case $B>0$ only.

Possible solutions $x=x\left( \eta \right) $ of the equation (\ref{abe501})
are functions of the parameter $\eta $ (of $\mu ,$ $\gamma /M^{2}$, $\Theta $%
) and are labelled by $m=0,1,...$ . One can find the following asymptotic
representations for these solutions at $\left| \eta \right| \rightarrow 0$%
\thinspace , 
\begin{eqnarray}
&&x_{m}\left( \eta \right) =m+\Delta x_{m},\;\;\Delta x_{m}=\frac{\sin
\left( \pi \mu \right) \Gamma (m+1-\mu )}{\pi \Gamma (m)Q\left( m\right) }%
\eta \,,\;\;m=1,2,3,\ldots ,  \notag \\
&&_{-}x_{0}\left( \eta \right) =-\frac{\eta M^{2}}{\gamma \Gamma \left( \mu
\right) }\;.  \label{abe502}
\end{eqnarray}
All $x_{m}\left( 0\right) ,\;m=1,2,...$ are positive and integer. The
asymptotic representation of $_{+}x_{0}\left( \eta \right) $ at $\left| \eta
\right| \rightarrow 0$ is discussed below. The function $_{+}x_{0}\left(
\eta \right) $ vanishes at the point $\eta =2\Gamma (\mu )$ and, in the
neighborhood of the latter point, has the form 
\begin{equation}
\,_{+}x_{0}\left( \eta \right) =\frac{\Gamma (\mu )-\eta /2}{\Gamma (\mu
)\left( \psi (\mu )-\psi (1)\right) }\,.  \label{abe504}
\end{equation}
Here $\psi (x)$ is the logarithmic derivative of the gamma function $\Gamma
(x)$, and $-\psi (1)\simeq 0.577\,$ is the Euler-Mascheroni constant \cite
{HTF1}. At $\left| \eta \right| \rightarrow \infty $ we found the following
asymptotic representations, 
\begin{eqnarray}
&&_{\varsigma }x_{m}\left( \varsigma \eta \right) =m+\mu +\Delta
x_{m},\;\;m=0,1,2,\ldots \;,{}\;\eta \rightarrow \infty \,,  \notag \\
&&_{\varsigma }x_{m}\left( \varsigma \eta \right) =m-1+\mu +\Delta
x_{m},\;\;m=1,2,3,\ldots \;,{}\;\eta \rightarrow -\infty \,,  \notag \\
&&\Delta x_{m}=-\frac{\sin \left( \pi \mu \right) \Gamma (m+\mu )Q\left(
m+\mu \right) }{\pi \Gamma (m+1)\eta }\;.  \label{abe505}
\end{eqnarray}
These approximations hold true only for $\left| \Delta x_{m}\right| $ $\ll
\mu \,$, $\left| x_{0}\left( \eta \right) \right| \ll \mu $.

According to\ \cite{W80} (see there Theorem 8.19, Corollary 1) if $T_{1}$
and $T_{2}$ are two self-adjoint extensions of the same symmetric operator
with equal finite defect indices $(d,d)$ then any interval $\left(
a,b\right) \subset \mathcal{R}$ not intersecting the spectrum of $T_{1}$
contains only isolated eigenvalues of the operator $T_{2}$ with total
multiplicity at most $d$. Let us select the extension $h^{\Omega }$ at $%
\Theta =\pi /2$ with the eigenvalues $\,_{+}\varepsilon =M\sqrt{1+2\gamma
\,_{+}x_{0}\left( \infty \right) /M^{2}}$\ and $\,_{\pm }\varepsilon =\pm M%
\sqrt{1+2\gamma \,_{\pm }x_{m}\left( \pm \infty \right) /M^{2}}$, $m\geq 1$.
Then the above theorem implies that if $\left( a,b\right) $ is an open
interval where$\ a,\,b$ are two subsequent eigenvalues of $h^{\Omega }$ at $%
\Theta =\pi /2$, or $_{\pm }\varepsilon =0$, then any self-adjoint extension 
$h^{\Omega }$ at $\Theta \neq \pi /2$\ has at most one eigenvalue in $\left(
a,b\right) $. According to\ \cite{AG81} (see there Chapter VIII Sect. 105
Theorem 3) for any $\varepsilon \in \left( a,b\right) $ there exist a
self-adjoint extension $h^{\Omega }$ with the eigenvalue $\varepsilon $. As
it follows from (\ref{abe501}), (\ref{abe505}), on the ranges $(m-1+\mu \leq
\,_{\pm }x_{m}\left( \eta \right) \leq m+\mu $, $m\geq 1)$, $(-M^{2}/2\gamma
\leq \,_{+}x_{0}\left( \eta \right) \leq \mu )$ the functions $_{\pm
}x\left( \eta \right) =\left( \,_{\pm }\varepsilon ^{2}-M^{2}\right)
/2\gamma $ are one-valued and continuous. This observation is in complete
agreement with the above general Theorems. The functions $\,_{\pm
}x_{m}\left( \eta \right) $ were found numerically in the weak field, $%
\gamma /M^{2}\ll 1$, for some first $m$'s. The plots of these functions (for 
$\mu =0.8$) see on Figs. 1 and 2.

\begin{figure} [!ht]
\centering
\label{kkk}
\includegraphics[width=4 in]{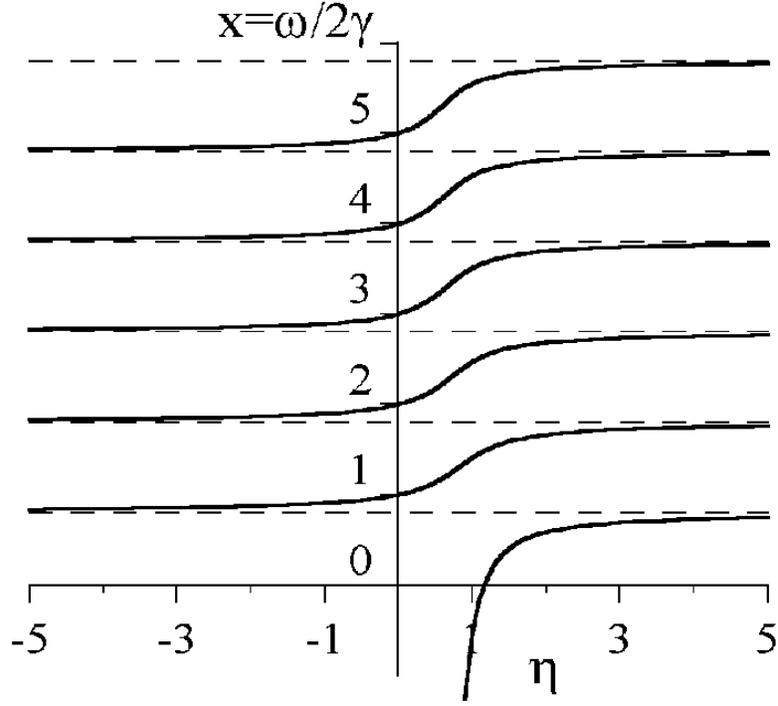}
\caption{Particle lowest energy levels in dependence on the parameter $%
\protect\eta _{+}=\frac{\Gamma \left( \protect\mu \right) }{\Gamma \left( 1-%
\protect\mu \right) }\left( \frac{\protect\gamma }{2M^{2}}\right) ^{1-%
\protect\mu }\tan \left( \frac{\protect\pi }{4}+\frac{\Theta
}{2}\right) $ }
\end{figure}

\begin{figure}[!ht]
\centering 
\includegraphics[width=4 in]{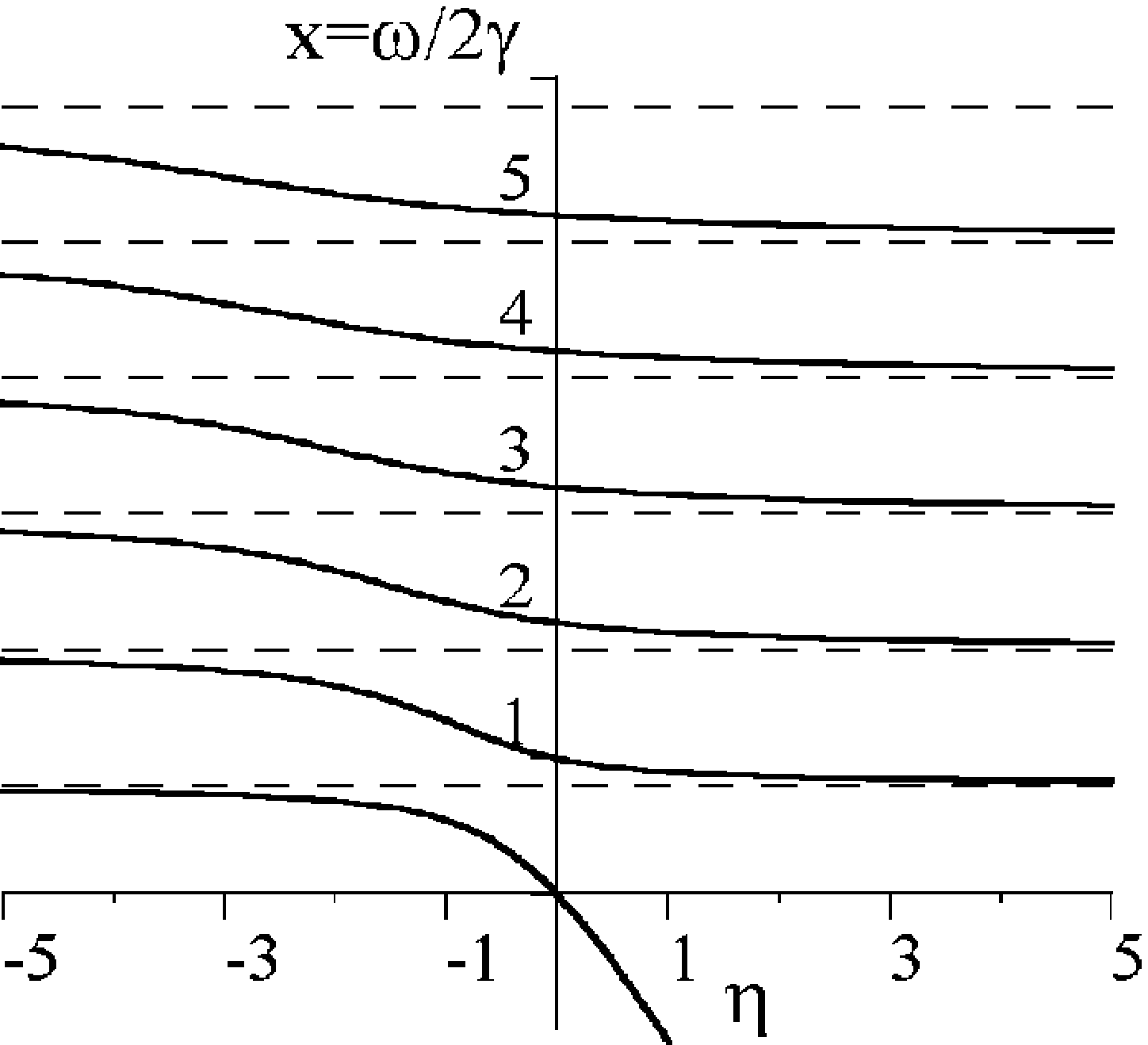}
\caption{Antiparticle lowest energy levels in dependence on the parameter $%
\protect\eta _{-}=\frac{\Gamma \left( \protect\mu \right) }{\Gamma \left( 1-%
\protect\mu \right) }\left( \frac{\protect\gamma }{2M^{2}}\right) ^{-\protect%
\mu }\tan \left( \frac{\protect\pi }{4}+\frac{\Theta }{2}\right) $}
\end{figure}

One can see that $\delta x_{m}=x_{m+1}\left( \eta \right) -x_{m}\left( \eta
\right) \rightarrow 1$ with increasing $m$. It follows from the equation (%
\ref{abe501}) that 
\begin{equation}
\delta x_{m}-1=\pi ^{-1}\left\{ \cot \left( \pi x_{m}\right) -\cot \left[
\pi \left( x_{m}-\mu \right) \right] \right\} ^{-1}\left( \frac{1-\mu }{x_{m}%
}-\delta Q\right) ,\;m\gg 1\,,  \label{abe506}
\end{equation}
where $\delta Q=\left. \frac{d}{dx}\ln Q(x)\right| _{x=x_{m}}\leq 1/x_{m}\,$%
. The curve $x_{5}\left( \eta \right) $ may give an idea how the functions $%
x_{m}\left( \eta \right) $ behave at big $m\,$.

Below we discuss some limiting cases.

Consider weak fields $B$, for which $\gamma /M^{2}\ll 1$, and
nonrelativistic electron energies, $x_{m}\left( \eta \right) \gamma
/M^{2}\ll 1$. Here the functions $\,_{\pm }x\left( \eta \right) $ change
significantly in the neighborhood of $\eta =0$ only. Beyond the neighborhood
of $\eta =0$ the functions $\,_{\pm }x\left( \eta \right) $ take the values
close to the corresponding asymptotic values given by (\ref{abe505}).

In the ultrarelativistic case, $x_{m}\left( \eta \right) \gamma /M^{2}\gg 1$%
, the behavior of $x_{m}\left( \eta \right) $ qualitatively depends on $\mu
\,$. One can distinguish three cases: $\mu <1/2$, $\mu >1/2$, $\mu =1/2$. If 
$\mu <1/2$ then the interval near $\eta =0$ on which the functions change
significantly diminishes with $m$ increasing. If $\mu >1/2$ then this
interval grows with $m$ increasing. For $\mu =1/2$ and $-\frac{1}{2}<\left( 
\frac{1}{4}+\frac{\Theta }{2\pi }\right) <\frac{1}{2}$, we get the
asymptotic representation, 
\begin{equation}
\,_{\varsigma }x_{m}\left( \eta \right) =m+\varsigma \left( \frac{1}{4}+%
\frac{\Theta }{2\pi }\right) ,{}\;m\gg 1\;.  \label{abe506a}
\end{equation}

One can see that negative $_{\pm }x_{0}\left( \eta \right) $ exist only for $%
\eta >0$ . That is, in the problem under consideration for $\pi /2<\Theta
<3\pi /2$ there exist only one particle state and only one antiparticle
state with energies $\left| \varepsilon \right| <M$. The same situation was
observed in the pure AB field case \cite{G89}. The minimal admissible
negative $x_{0}\left( \eta \right) $ is defined by the condition $%
\varepsilon =0$. In strong fields $B$, for which $\gamma /M^{2}\sim 1$, the
quantity $x_{0}\left( \eta \right) $\ is close to zero. Let $\Theta _{0}$
correspond to such an extension that admits $\varepsilon =0$. The value of $%
\Theta _{0}$ is defined by the expression 
\begin{equation}
\tan \left( \frac{\pi }{4}+\frac{\Theta _{0}}{2}\right) =-\frac{\Gamma
(1-\mu )\Gamma (\mu +M^{2}/2\gamma )}{2^{\mu }\Gamma (\mu )\Gamma
(1+M^{2}/2\gamma )}\left( \frac{M^{2}}{\gamma }\right) ^{1-\mu }.
\label{abe507}
\end{equation}
In weak fields, $\gamma /M^{2}\ll 1$, $x_{0}\left( \eta \right) $ take big
absolute values, and the angle $\Theta _{0}$ is defined by the expression 
\begin{equation}
\tan \left( \frac{\pi }{4}+\frac{\Theta _{0}}{2}\right) =-\frac{\Gamma
(1-\mu )}{2^{2\mu -1}\Gamma (\mu )}\,,  \label{abe508}
\end{equation}
and does not depend on the magnetic field. It follows from (\ref{abe507})
that in the superstrong fields $B$, for which $\gamma /M^{2}\gg 1$, the
angle $\Theta _{0}$ does not depend on the magnetic field as well.

In weak magnetic fields, $\gamma /M^{2}\ll 1$, and for nonrelativistic
energy values, $x_{0}\gamma /M^{2}\ll 1$, one can get relations 
\begin{eqnarray}
&&\,_{+}x_{0}\left( \eta \right) =-\left( 2/\eta \right) ^{1/\left( 1-\mu
\right) }\,,  \label{abe509} \\
&&\,_{-}x_{0}\left( \eta \right) =-\left( \eta M^{2}/\gamma \right) ^{1/\mu
}\,  \label{abe5010}
\end{eqnarray}
that are valid when $\eta $\ is small in (\ref{abe509}) and $\eta
M^{2}/\gamma \gg 1$ in (\ref{abe5010}).

Let us consider the particular case $\Theta =-\pi /2.$ It follows from (\ref
{abe48}) that for$\ B>0$, there exists $_{-}\varepsilon =-M$. The energies $%
\left| \varepsilon \right| >M$ are defined by poles of $\Gamma (1-x)$ or of $%
\Gamma (1-\mu -x)$ for $B>0$ or $B<0$, respectively. The spectrum $%
\varepsilon $ coincides with one defined by Eqs. (\ref{abe30}), (\ref{abe31}%
) for $\psi ^{I}$. Moreover, using the relation (\ref{4.140}), we can see
that the spinors $\psi _{\omega }(r)$ coincide with $\psi ^{I}$ up to a
normalization constant, 
\begin{equation}
\psi _{\omega }\left( r\right) \propto \psi ^{I}\left( r\right) \;\mathrm{%
for\;}\Theta =-\pi /2\,.  \label{abe50a}
\end{equation}

In the case $\Theta =\pi /2$ we have the following picture: It follows from (%
\ref{abe48}) that for $B<0$ there exists $_{+}\varepsilon =M$. The energies $%
\left| \varepsilon \right| >M$ are defined by poles of $\Gamma (\mu -x)$ or
of $\Gamma (1-x)$ for $B>0$ or $B<0$, respectively. The spectrum $%
\varepsilon $ coincides with one found by Eqs. (\ref{abe30}), (\ref{abe31})\
for $\psi ^{II}$. From (\ref{4.140}) it follows that the spinor $\psi
_{\omega }(r)$ coincides with $\psi ^{II}$ up to a normalization constant, 
\begin{equation}
\psi _{\omega }(r)\propto \psi ^{II}\left( r\right) \;\mathrm{for\;}\Theta
=\pi /2\,.  \label{abe50b}
\end{equation}

Using results for $B<0$ which are presented in Appendix B one can conclude
that the spectrum asymmetry takes place for the spinning particles in the
magnetic-solenoid field. There is a relation between the three-dimensional
chiral anomaly and fermion zero modes in a uniform magnetic field\ \cite
{NS83} (for review see \cite{J82,NS86}). One can see the effect also takes
place in the AB potential presence.

The spectrum asymmetry is known in 2+1 QED for the uniform magnetic field.
In the uniform magnetic field the states with $\omega =0$ for $l\neq 0$ are
observed if \textrm{sgn}$l=-\mathrm{sgn}B$ (for antiparticle if $B>0$ and
for particle if $B<0$). The spectrum changes mirror-like with the change of
the magnetic field sign. One can see that for $l\neq 0$ the spectrum
properties in the magnetic-solenoid field is similar to the spectrum
properties in the uniform magnetic field. The presence of the AB potential
is especially essential for the states with $l=0$, when the particle
penetrates the solenoid.

Spectra in $3+1$ dimensions can be obtained from the results in $2+1$
dimensions. Namely, we use the fact that the solutions $\psi _{\varepsilon
,1}(x_{\perp })$ in $3+1$ dimensions are obtained from the solutions $\psi
_{\varepsilon }^{(1)}(x_{\perp })$ in $2+1$ dimensions. Thus, spectra in $%
3+1 $ dimensions can be obtained from the results in $2+1$ dimensions with
the substitution $M$ by $\widetilde{M}$, and the relation (\ref{abe5b}). As
a consequence, we obtain an additional interpretation of Figs. 1, 2. In
particular, Fig.1 presents energy lowest levels for particles with spin $s=1$%
, and Fig. 2 presents energy lowest levels for particles with spin $s=-1$.

\section{Solenoid regularization}

One can introduce the AB field as a limiting case of a finite radius
solenoid field (the regularized AB field). In this way, one can fix the
extension parameters. First, the manner of doing that in the pure AB field
was presented by Hagen \cite{H90}. Below, we consider the problem in the
presence of the uniform magnetic field. To this end we have to study
solutions of the Dirac equation (\ref{abe1}) in the combination of the
regularized AB field and the uniform magnetic field.

Let the solenoid have a radius $R$. We assume that inside the solenoid there
is an axially symmetric magnetic field $B^{in}(r)$ that creates the flux $%
\Phi =\left( l_{0}+\mu \right) \Phi _{0},\;\Phi _{0}=2\pi /e$. Outside the
solenoid ($r>R$) the field $B^{in}(r)$ vanishes. Thus, $e\int_{0}^{R}B^{in}%
\left( r\right) rdr=l_{0}+\mu $. The function $B^{in}(r)$ is arbitrary but
such that integrals in the functions $\vartheta \left( x\right) ,\;b\left(
x\right) $ in (\ref{sol-inside}) are not divergent. We select the potentials
of the field $B^{in}(r)$ in the form 
\begin{equation}
eA_{1}^{in}=\vartheta \left( x\right) \frac{\sin \varphi }{Rx}%
,\;eA_{2}=-\vartheta \left( x\right) \frac{\cos \varphi }{Rx}\,,
\label{abe52}
\end{equation}
where 
\begin{equation*}
\;\vartheta \left( x\right) =\int_{0}^{x}f\left( x^{\prime }\right)
x^{\prime }dx^{\prime },\;f(x)=R^{2}eB^{in}(xR),\;x=r/R\;.
\end{equation*}
The potentials of the uniform magnetic field are 
\begin{equation}
A_{0}=0,\;A_{1}=A\left( r\right) \frac{\sin \varphi }{r},\;A_{2}=-A\left(
r\right) \frac{\cos \varphi }{r}\;,\;\;A\left( r\right) =Br^{2}/2\;.
\label{x.1}
\end{equation}
Outside the solenoid the potentials have the form (\ref{abe2}).

Let us analyze solutions of the Dirac equation in the above defined field.
To this end we have to solve the equation inside and outside the solenoid
and continuously join the corresponding solutions. The former Dirac spinors
we are going to call the inside solutions, whereas the latter ones the
outside solutions.

First, we study the problem in $2+1$ dimensions. We demand the solutions to
be square integrable and regular at $r\rightarrow 0$. By the same manner as
in the Sect. 2, we can find that the inside radial spinors $\psi _{\omega
,l}^{in}(r)$\ ($r\leq R$) obey the equation: 
\begin{equation*}
h^{in}\psi _{\omega ,l}^{in}\left( r\right) =\varepsilon \psi _{\omega
,l}^{in}\left( r\right) ,\;\;h^{in}=\Pi ^{in}+\sigma ^{3}M\;,
\end{equation*}
where 
\begin{equation}
\Pi ^{in}=-\frac{i}{R}\left\{ \partial _{x}+\frac{\sigma ^{3}}{x}\left[
l-l_{0}-\frac{1}{2}\left( 1-\sigma ^{3}\right) +\vartheta \left( x\right)
+\xi \rho _{R}x^{2}\right] \right\} \sigma ^{1},\;\rho _{R}=\gamma R^{2}/2\;.
\label{abe55}
\end{equation}
We demand the functions $\psi _{\omega ,l}^{in}\left( r\right) $ to be
square integrable on the interval $\left( 0,R\right) $. For $\omega =0$ ($%
\left| \varepsilon \right| =M$) the solutions read, 
\begin{eqnarray}
_{+}\psi _{0,l}^{in}(r) &=&\phi _{0,l,1}^{in}(x)\upsilon
_{1}\;,{}\;l-l_{0}\geq 1\;,  \notag \\
_{-}\psi _{0,l}^{in}(r) &=&\phi _{0,l,-1}^{in}(x)\upsilon
_{-1}\;,{}\;l-l_{0}\leq 0\;,  \notag \\
\phi _{0,l,\sigma }^{in}(x) &=&cx^{|\eta |}\exp \left\{ \sigma \int_{0}^{x}d%
\tilde{x}\tilde{x}^{-1}\left( \vartheta (\tilde{x})+\xi \rho _{R}\tilde{x}%
^{2}\right) \right\} ,\;\eta =l-l_{0}-\left( 1+\sigma \right) /2\;,
\label{abe56}
\end{eqnarray}
where $c$ is an arbitrary constant. For $\omega \neq 0$ we present the
spinors in the form 
\begin{equation*}
\psi _{\omega ,l}^{in}\left( r\right) =\left( 
\begin{array}{c}
\psi _{1}^{in}(r) \\ 
\psi _{2}^{in}(r)
\end{array}
\right) =\left[ \sigma ^{3}\left( \varepsilon -\Pi ^{in}\right) +M\right] 
\left[ c_{1}\phi _{l,1}^{in}(x)\upsilon _{1}+ic_{-1}\phi
_{l,-1}^{in}(x)\upsilon _{-1}\right] \,,
\end{equation*}
where $c_{\sigma }$ are arbitrary constants. The functions $\phi _{l,\sigma
}^{in}\left( x\right) $ satisfy the equation 
\begin{equation}
\left[ \frac{1}{x}\frac{\partial }{\partial x}x\frac{\partial }{\partial x}-%
\frac{1}{x^{2}}\left( \eta +\vartheta \left( x\right) +\xi \rho
_{R}x^{2}\right) ^{2}+\omega R^{2}-\sigma \left( f\left( x\right) +2\xi \rho
_{R}\right) \right] \phi _{l,\sigma }^{in}\left( x\right) =0  \label{abe58}
\end{equation}
and must be regular at $r=0$ in order to satisfy the square integrability
condition for $\psi _{\omega ,l}^{in}\left( r\right) $. We are interested in
the limiting case $R\rightarrow 0$. For our purposes it is enough to use the
approximation $\rho _{R}\ll 1$, $\omega R^{2}\ll 1$. Dropping terms
proportional to $R^{2}$ in (\ref{abe55}) and (\ref{abe58}), we find that
solutions of Eq. (\ref{abe58}) have the form 
\begin{eqnarray}
&&\phi _{l,\sigma }^{in}\left( x\right) =\left\{ 
\begin{array}{c}
cx^{|\eta |}e^{\sigma b(x)},\;\sigma \eta \geq 0\,, \\ 
cx^{-|\eta |}e^{\sigma b(x)}\int_{0}^{x}d\tilde{x}\tilde{x}^{2|\eta
|-1}e^{-2\sigma b(\tilde{x})},\;\sigma \eta <0\,,
\end{array}
\right.  \notag \\
&&b(x)=\int_{0}^{x}d\tilde{x}\tilde{x}^{-1}\vartheta (\tilde{x})\,.
\label{sol-inside}
\end{eqnarray}

The outside solutions ($r\geq R$) obey the equation 
\begin{equation}
h\psi _{\omega ,l}^{out}\left( r\right) =\varepsilon \psi _{\omega
,l}^{out}\left( r\right)  \label{abe57}
\end{equation}
and must be square integrable on the interval $\left( R,\infty \right) $.
Here $h$ is defined by Eqs. (\ref{abe12}), (\ref{abe13}). The general form
of the outside solutions reads: 
\begin{eqnarray}
&&\psi _{\omega ,l}^{out}\left( r\right) =\left[ \sigma ^{3}\left(
\varepsilon -\Pi \right) +M\right] \left( c_{1}\phi _{l,1}^{out}(r)\upsilon
_{1}+ic_{-1}\phi _{l,-1}^{out}(r)\upsilon _{-1}\right) \,,  \notag \\
&&\phi _{l,\sigma }^{out}\left( r\right) =\psi _{\lambda ,\alpha }\left(
\rho \right) ,\;\alpha =l+\mu -\left( 1+\sigma \right) /2,\;2\lambda =\omega
/\gamma -\xi \left( l+\mu -\left( 1-\sigma \right) /2\right) \,.
\label{s-out}
\end{eqnarray}
The solutions $\psi _{\omega ,l}^{out}\left( r\right) $ and $\psi _{\omega
,l}^{in}\left( r\right) $ must be joined continuously at $r=R$, 
\begin{equation}
\psi ^{out}\left( R\right) =\psi ^{in}\left( R\right)  \label{bc-out-in}
\end{equation}
and obey the normalization relation 
\begin{eqnarray}
&&N_{\omega ,l}^{in}+N_{\omega ,l}^{out}=1\,,  \notag \\
N_{\omega ,l}^{in} &=&\int_{0}^{R}\left( \psi _{\omega ,l}^{in}(r)\right)
^{\dagger }\psi _{\omega ,l}^{in}(r)rdr\;,\;\;N_{\omega
,l}^{out}=\int_{R}^{\infty }\left( \psi _{\omega ,l}^{out}(r)\right)
^{\dagger }\psi _{\omega ,l}^{out}(r)rdr\,.  \label{abe611}
\end{eqnarray}
One can treat the AB field as a limiting case of a finite radius solenoid
field if 
\begin{equation}
\lim_{\rho _{R}\rightarrow 0}N_{\omega ,l}^{in}=0\,.  \label{x.2}
\end{equation}
The joining condition (\ref{bc-out-in}) one can realize using the following
conditions for the functions $\phi _{l,\sigma }^{in}\left( r\right) $\ and $%
\phi _{l,\sigma }^{out}\left( r\right) $ at $r=R$, 
\begin{equation}
\phi \left( R-\epsilon \right) =\phi \left( R+\epsilon \right) ,\;\frac{d}{dr%
}\phi \left( R-\epsilon \right) =\frac{d}{dr}\phi \left( R+\epsilon \right)
\,.  \label{cond-cont}
\end{equation}

It is convenient to use in (\ref{s-out}) the representation (\ref{4.140})
for $\psi _{\lambda ,\alpha }\left( \rho \right) $. Then, the functions $%
\phi _{l,\sigma }^{out}\left( r\right) $ read: 
\begin{eqnarray}
&&\phi _{l,\sigma }^{out}\left( r\right) =a_{\sigma }I_{n_{\sigma
},m_{\sigma }}\left( \rho \right) +b_{\sigma }I_{m_{\sigma },n_{\sigma
}}\left( \rho \right) ,\;n_{\sigma }=\lambda -\frac{1-\alpha }{2}%
,\;m_{\sigma }=\lambda -\frac{1+\alpha }{2}\,,  \notag \\
&&a_{\sigma }=K\sin n_{\sigma }\pi ,\;b_{\sigma }=-K\sin m_{\sigma }\pi ,\;K=%
\frac{\sqrt{\Gamma \left( 1+n_{\sigma }\right) \Gamma \left( 1+m_{\sigma
}\right) }}{\sin \left( n_{\sigma }-m_{\sigma }\right) \pi }\,.
\label{abe59}
\end{eqnarray}
where $n_{\sigma }$, $m_{\sigma }$ are real numbers.

Using (\ref{cond-cont}) one can find the coefficients $a_{\sigma
},\;b_{\sigma }$: for the case $l-l_{0}\leq 0$, 
\begin{eqnarray}
&&a_{1}=\rho _{R}^{-\left( l+\mu -1\right) /2}c\tilde{a}_{1},{}\;b_{1}=\rho
_{R}^{\left( l+\mu -1\right) /2}c\tilde{b}_{1},  \label{abe60a} \\
&&a_{-1}=\rho _{R}^{-\left( l+\mu \right) /2+1}c\tilde{a}_{-1},{}\;b_{-1}=%
\rho _{R}^{\left( l+\mu \right) /2}c\tilde{b}_{-1},  \label{abe60b}
\end{eqnarray}
whereas for the case $l-l_{0}>0$, 
\begin{eqnarray}
&&a_{1}=\rho _{R}^{-\left( l+\mu -1\right) /2}c^{\prime }\tilde{a}%
_{1}^{\prime },{}\;b_{1}=\rho _{R}^{\left( l+\mu -1\right) /2+1}c^{\prime }%
\tilde{b}_{1}^{\prime },  \label{abe61a} \\
&&a_{-1}=\rho _{R}^{-\left( l+\mu \right) /2}c^{\prime }\tilde{a}%
_{-1}^{\prime },{}\;b_{-1}=\rho _{R}^{\left( l+\mu \right) /2}c^{\prime }%
\tilde{b}_{-1}^{\prime },  \label{abe61b}
\end{eqnarray}
where the non-vanishing coefficients $\tilde{a}$, $\tilde{b}$, $\tilde{a}%
^{\prime }$, $\tilde{b}^{\prime }$ are not depending on $\rho _{R}$, the
coefficients $c$, $c^{\prime }$ are normalizing factors which depend on $%
\rho _{R}$.

Calculating the normalization factors one obtains at $R\rightarrow 0$, for $%
l-l_{0}\leq 0$, 
\begin{eqnarray*}
a_{1} &=&const\neq 0,\;b_{1}=0,\;a_{-1}=\;b_{-1}=0,\;l\geq 1\,, \\
a_{1} &=&0,\;b_{1}=const\neq 0,\;a_{-1}=\;b_{-1}=0,\;l\leq 0\,,
\end{eqnarray*}
whereas for $l-l_{0}>0$, 
\begin{eqnarray*}
a_{1} &=&b_{1}=0,\;a_{-1}=const\neq 0,\;b_{-1}=0,\;l\geq 0\,, \\
a_{1} &=&b_{1}=0,\;a_{-1}=0,\;b_{-1}=const\neq 0,\;l\leq -1\,.
\end{eqnarray*}
For $l=0$ the value of the coefficients is defined by \textrm{sgn}$\Phi $.
One can verify that the condition (\ref{x.2}) is satisfied.

Thus, one obtains that for any sign of $B$ the solutions are expressed via
Laguerre polynomials (\ref{abe31}). Particularly, for $l=0$ we find that\
the solutions $\psi _{\omega ,0}^{out}\left( r\right) $ coincide with either 
$\psi _{m}^{I}\left( r\right) $ or $\psi _{m}^{II}\left( r\right) $
accordingly to $\mathrm{sgn}\left( \Phi \right) $, 
\begin{equation}
\psi _{\omega ,0}^{out}\left( r\right) =\left\{ 
\begin{array}{c}
\psi _{m}^{I}\left( r\right) ,\;\mathrm{sgn}\left( \Phi \right) =+1 \\ 
\psi _{m}^{II}\left( r\right) ,\;\mathrm{sgn}\left( \Phi \right) =-1
\end{array}
\right. \,.  \label{abe62}
\end{equation}

In Sect. 3 we have found the relation between the extension parameter values
and solution types in the critical subspace $l=0$ (\ref{abe50a}), (\ref
{abe50b}). Now we are in position to refine this relation. Namely, if one
introduces the AB field as a field of the finite radius solenoid for a
zero-radius limit, then the extension parameter $\Theta $ is fixed to be $%
\Theta =-\mathrm{sgn}\left( \Phi \right) \pi /2$. Besides, this way of the
AB field introduction explicitly implies no additional interaction in the
solenoid core.

To solve the problem in $3+1$ dimensions we use the results in $2+1$
dimensions presented above. In the limit $R\rightarrow 0$\ solutions in the
critical subspace have the form 
\begin{equation}
\Psi _{s}^{out}\left( x_{\bot }\right) =N\left( 
\begin{array}{c}
\left[ 1+\left( p^{3}+s\widetilde{M}\right) /M\right] g_{0}\left( \varphi
\right) \psi _{\omega ,l}^{out}\left( r\right) \\ 
\left[ -1+\left( p^{3}+s\widetilde{M}\right) /M\right] g_{0}\left( \varphi
\right) \psi _{\omega ,l}^{out}\left( r\right)
\end{array}
\right) \,,  \label{r26}
\end{equation}
where the functions $g_{0}\left( \varphi \right) $, $\psi _{\omega
,l}^{out}\left( r\right) $ are defined in (\ref{abe11}) and\ (\ref{abe62}), (%
\ref{abe31}), respectively. We specify the values of the extension
parameters in $3+1$ dimensions as follows, 
\begin{equation}
\Theta _{+1}=\Theta _{-1}=-\mathrm{sgn}\left( \Phi \right) \pi /2\,.
\label{r27}
\end{equation}

\section{Summary}

We have studied in detail solutions of the Dirac equation in the
magnetic-solenoid field in $2+1$ and $3+1$ dimensions. In the general case,
solutions in $2+1$ and $3+1$ dimensions are not related in a simple manner.
However, it has been demonstrated that solutions in $3+1$ dimensions with
special spin quantum numbers can be constructed directly on the base of
solutions in $2+1$ dimensions. To this end, one has to choose the $z$%
-component of the polarization pseudovector $S^{3}$ as the spin operator in $%
3+1$ dimensions. This is a new result not only for the magnetic-solenoid
field background, but for the pure AB field as well. The choice $S^{3}$ as
the spin operator was convenient from different points of view. For example,
solutions with arbitrary momentum $p^{3}$ are eigenvectors of the operator $%
S^{3}$. This allows us to separate explicitly spin and coordinate variables
in $3+1$ dimensions. Thus, in $3+1$ dimensions one has to study self-adjoint
extensions of the radial Hamiltonian only.\ Moreover, boundary conditions in
such a representation do not violate translation invariance along the
natural direction which is the magnetic-solenoid field direction. The
self-adjoint extensions of the Dirac Hamiltonian in the magnetic-solenoid
field have been constructed using von Neumann's theory of deficiency
indices. A one-parameter family of allowed boundary conditions in $2+1$
dimensions and a two-parameter family in $3+1$ dimensions have been
constructed. By that the complete orthonormal sets of solutions have been
found. The energy spectra dependent on the extension parameter $\Theta $
have been defined for the different self-adjoint extensions. Besides, for
the first time solutions of the Dirac equation in the regularized
magnetic-solenoid field have been described in detail. We considered an
arbitrary magnetic field distribution inside a finite-radius solenoid. It
was shown that the extension parameters $\Theta =-\mathrm{sgn}(\Phi )\pi /2$
in $2+1$ dimensions and $\Theta _{+1}=\Theta _{-1}=-\mathrm{sgn}(\Phi )\pi
/2 $ in $3+1$ dimensions correspond to the limiting case $R\rightarrow 0$ of
the regularized magnetic-solenoid field.

\section{Acknowledgments}

D.M.G. thanks CNPq and FAPESP for permanent support. A.A.S. thanks FAPESP
for support. S.P.G. acknowledges the support of CAPES and FAPESP, grant
02/11321-8. S.P.G. and A.A.S. thank the Department of Physics of
Universidade Federal de Sergipe (Brazil) for hospitality. We thank Prof.
I.V. Tyutin for useful discussions.

\setcounter{section}{0} \renewcommand{\thesection}{\Alph{section}}

\section{Appendix}

1.The Laguerre function $I_{n,m}(x)$ is defined by the relation

\begin{equation}
I_{n,m}(x)=\sqrt{\frac{\Gamma \left( 1+n\right) }{\Gamma \left( 1+m\right) }}%
\frac{\exp \left( -x/2\right) }{\Gamma \left( 1+n-m\right) }x^{\left(
n-m\right) /2}\Phi (-m,n-m+1;x)\,.  \label{4.1}
\end{equation}
Here $\Phi \left( a,b;x\right) $ is the confluent hypergeometric function in
a standard definition (see \cite{GR94}, 9.210). Let $m$ be a non-negative
integer number; then the Laguerre function is related to Laguerre
polynomials $L_{m}^{\alpha }(x)$ (\cite{GR94}, 8.970, 8.972.1) by the
equation

\begin{eqnarray}
I_{m+\alpha ,m}(x) &=&\sqrt{\frac{m!}{\Gamma \left( m+\alpha +1\right) }}%
e^{-x/2}x^{\alpha /2}L_{m}^{\alpha }(x)\,,  \label{4.42} \\
\;L_{m}^{\alpha }(x) &=&\frac{1}{m!}e^{x}x^{-\alpha }\frac{d^{m}}{dx^{m}}%
e^{-x}x^{m+\alpha }\,.  \label{4.5}
\end{eqnarray}
Using well-known properties of the confluent hypergeometric function ( \cite
{GR94}, 9.212; 9.213; 9.216), one can easily get the following relations for
the Laguerre functions

\begin{eqnarray}
&&2\sqrt{x(n+1)}I_{n+1,m}(x)=(n-m+x)I_{n,m}(x)-2xI_{n,m}^{\prime }(x)\;,
\label{4.11} \\
&&2\sqrt{x(m+1)}I_{n,m+1}(x)=(n-m-x)I_{n,m}(x)+2xI_{n,m}^{\prime }(x)\;,
\label{4.12} \\
&&2\sqrt{xn}I_{n-1,m}(x)=(n-m+x)I_{n,m}(x)+2xI_{n,m}^{\prime }(x)\;,
\label{4.13} \\
&&2\sqrt{xm}I_{n,m-1}(x)=(n-m-x)I_{n,m}(x)-2xI_{n,m}^{\prime }(x).
\label{4.16}
\end{eqnarray}
Using properties of the confluent hypergeometric function, one can get a
representation

\begin{equation}
I_{n,m}(x)=\sqrt{\frac{{\Gamma (}1+n{)}}{{\Gamma (}1+m{)}}}{\frac{{\exp }%
\left( x/2\right) }{{\Gamma (}1+n-m{)}}}x^{\frac{{n-m}}{2}}\Phi
(1+n,1+n-m;-x)\;,  \label{4.24}
\end{equation}
and a relation (\cite{GR94}, 9.214)

\begin{equation}
I_{n,m}(x)=(-1)^{n-m}I_{m,n}(x),\;n-m\;\mathrm{integer}\;.  \label{4.25}
\end{equation}
The functions $I_{\alpha +m,m}(x)$ obey the orthonormality relation

\begin{equation}
\int_{0}^{\infty }I_{\alpha +n,n}\left( x\right) I_{\alpha +m,m}\left(
x\right) dx=\delta _{m,n}\;,  \label{4.41}
\end{equation}
which follows from the corresponding properties of the Laguerre polynomials
( \cite{GR94}, 7.414.3). The set of the Laguerre functions 
\begin{equation*}
I_{\alpha +m,m}(x),\;m=0,1,2...\,\,,\;\alpha >-1
\end{equation*}
is complete in the space of square integrable functions on the half-line ($%
x\geq 0$), 
\begin{equation}
\sum_{m=0}^{\infty }I_{\alpha +m,m}(x)I_{\alpha +m,m}(y)=\delta \left(
x-y\right) \,.  \label{ap18}
\end{equation}

2. The function $\psi _{\lambda ,\alpha }(x)$ is even with respect to index $%
\alpha $, 
\begin{equation}
\psi _{\lambda ,\alpha }\left( x\right) =\psi _{\lambda ,-\alpha }\left(
x\right) .  \label{4.138}
\end{equation}
It can be expressed via the confluent hypergeometric functions 
\begin{eqnarray}
&&\psi _{\lambda ,\alpha }\left( x\right) =e^{-\frac{x}{2}}\left[ \frac{%
\Gamma \left( -\alpha \right) x^{\frac{\alpha }{2}}}{\Gamma \left( \frac{%
1-\alpha }{2}-\lambda \right) }\Phi \left( \frac{1+\alpha }{2}-\lambda
,1+\alpha ;x\right) \right.  \notag \\
&&\left. +\frac{\Gamma \left( \alpha \right) x^{-\frac{\alpha }{2}}}{\Gamma
\left( \frac{1+\alpha }{2}-\lambda \right) }\Phi \left( \frac{1-\alpha }{2}%
-\lambda ,1-\alpha ;x\right) \right] ,  \label{4.139}
\end{eqnarray}
or, using (\ref{4.1}), via the Laguerre functions 
\begin{eqnarray}
&&\psi _{\lambda ,\alpha }\left( x\right) =\frac{\sqrt{\Gamma \left(
1+n\right) \Gamma \left( 1+m\right) }}{\sin \left( n-m\right) \pi }\left(
\sin n\pi I_{n,m}\left( x\right) -\sin m\pi I_{m,n}\left( x\right) \right) ,
\notag \\
&&\alpha =n-m,\,2\lambda =1+n+m,\,n=\lambda -\frac{1-\alpha }{2},\,m=\lambda
-\frac{1+\alpha }{2}.  \label{4.140}
\end{eqnarray}
There are the following relations of the functions $\psi _{\lambda ,\alpha
}\left( x\right) $,

\begin{eqnarray}
\psi _{\lambda ,\alpha }\left( x\right) &=&\sqrt{x}\psi _{\lambda -\frac{1}{2%
},\alpha -1}\left( x\right) +\frac{1+\alpha -2\lambda }{2}\psi _{\lambda
-1,\alpha }\left( x\right) ,  \notag \\
\psi _{\lambda ,\alpha }\left( x\right) &=&\sqrt{x}\psi _{\lambda -\frac{1}{2%
},\alpha +1}\left( x\right) +\frac{1-\alpha -2\lambda }{2}\psi _{\lambda
-1,\alpha }\left( x\right) ,  \notag \\
2x\psi _{\lambda ,\alpha }^{\prime }\left( x\right) &=&\left( 2\lambda
-1-x\right) \psi _{\lambda ,\alpha }\left( x\right) +\frac{1}{2}\left(
2\lambda -1-\alpha \right) \left( 2\lambda -1+\alpha \right) \psi _{\lambda
-1,\alpha }\left( x\right) ,  \notag \\
2x\psi _{\lambda ,\alpha }^{\prime }\left( x\right) &=&\left( \alpha
-x\right) \psi _{\lambda ,\alpha }\left( x\right) +\left( 2\lambda -1-\alpha
\right) \sqrt{x}\psi _{\lambda -\frac{1}{2},\alpha +1}\left( x\right)  \notag
\\
&=&\left( x-2\lambda -1\right) \psi _{\lambda ,\alpha }-2\psi _{\lambda
+1,\alpha }\,.  \label{4.144}
\end{eqnarray}
As a consequence of these properties we get 
\begin{eqnarray}
&&A_{\alpha }\psi _{\lambda ,\alpha }\left( x\right) =\frac{2\lambda
-1+\alpha }{2}\psi _{\lambda -\frac{1}{2},\alpha -1}\left( x\right)
,\,A_{\alpha }^{+}\psi _{\lambda -\frac{1}{2},\alpha -1}\left( x\right)
=\psi _{\lambda ,\alpha }\left( x\right) ,  \notag \\
&&A_{\alpha }=\frac{x+\alpha }{2\sqrt{x}}+\sqrt{x}\frac{d}{dx},\quad
A_{\alpha }^{+}=\frac{x+\alpha -1}{2\sqrt{x}}-\sqrt{x}\frac{d}{dx}.
\label{4.145}
\end{eqnarray}
Using well-known asymptotics of the Whittaker function (\cite{GR94}, 9.227),
we have 
\begin{equation}
\psi _{\lambda ,\alpha }\left( x\right) \sim x^{\lambda -\frac{1}{2}}e^{-%
\frac{x}{2}},\quad x\rightarrow \infty ;\;\;\psi _{\lambda ,\alpha }\left(
x\right) \sim \frac{\Gamma \left( |\alpha |\right) }{\Gamma \left( \frac{%
1+|\alpha |}{2}-\lambda \right) }x^{-\frac{|\alpha |}{2}},\,\,\alpha \neq
0,\quad x\sim 0\,.  \label{4.147}
\end{equation}
The function $\psi _{\lambda ,\alpha }\left( x\right) $ is correctly defined
and infinitely differentiable for $0<x<\infty $ and for any complex $\lambda
,\alpha .$ In this respect one can mention that the Laguerre function are
not defined for negative integer $n,m.$ In particular cases, when one of the
numbers $n,m$ is non-negative and integer, the function $\psi _{\lambda
,\alpha }\left( x\right) $ coincides (up to a constant factor) with the
Laguerre function.

According to (\ref{4.147}), the functions $\psi _{\lambda ,\alpha }\left(
x\right) $ are square integrable on the interval $0\leq x<\infty $ whenever $%
|\alpha |<1$. It is not true for $|\alpha |\geq 1.$ The corresponding
integrals at $\alpha \neq 0$ can be calculated as following (\cite{GR94},
7.611),

\begin{eqnarray}
&&\int\limits_{0}^{\infty }\psi _{\lambda ,\alpha }\left( x\right) \psi
_{\lambda ^{\prime },\alpha }\left( x\right) \,dx=\frac{\pi }{\left( \lambda
^{\prime }-\lambda \right) \sin \alpha \pi }\left\{ \left[ \Gamma \left( 
\frac{1+\alpha -2\lambda ^{\prime }}{2}\right) \Gamma \left( \frac{1-\alpha
-2\lambda }{2}\right) \right] ^{-1}\right.  \notag \\
&&\left. -\left[ \Gamma \left( \frac{1-\alpha -2\lambda ^{\prime }}{2}%
\right) \Gamma \left( \frac{1+\alpha -2\lambda }{2}\right) \right]
^{-1}\right\} ,\quad |\alpha |<1,  \label{4.148} \\
&&\int\limits_{0}^{\infty }|\psi _{\lambda ,\alpha }\left( x\right)
|^{2}\;dx=\frac{\pi }{\sin \alpha \pi }\frac{\psi \left( \frac{1+\alpha
-2\lambda }{2}\right) -\psi \left( \frac{1-\alpha -2\lambda }{2}\right) }{%
\Gamma \left( \frac{1+\alpha -2\lambda }{2}\right) \Gamma \left( \frac{%
1-\alpha -2\lambda }{2}\right) },\quad |\alpha |<1,  \label{4.150}
\end{eqnarray}
Here $\psi (x)$ is the logarithmic derivative of the $\Gamma $-function ( 
\cite{GR94}, 8.360). In the general case, the functions $\psi _{\lambda
,\alpha }\left( x\right) \;$and $\psi _{\lambda ^{\prime },\alpha }(x),$ $%
\lambda ^{\prime }\neq \lambda ,$ are not orthogonal, as it follows from (%
\ref{4.148}).

\section{Appendix}

Below we present modifications of some above formulas for the case $B<0$.

1. The spectrum of $\omega $ corresponding to the functions $\phi
_{m,l,\sigma }(r)$ is 
\begin{equation}
\omega =\left\{ 
\begin{array}{l}
2\gamma \left( m-l+1-\mu \right) ,\;{}l-\left( 1+\sigma \right) /2<0 \\ 
2\gamma \left( m+\left( 1-\sigma \right) /2\right) ,\;{}l-\left( 1+\sigma
\right) /2\geq 0
\end{array}
\right. ,  \label{abe23}
\end{equation}
and the spectrum of $\omega $ corresponding to the functions $\phi
_{m,\sigma }^{ir}(r)$ is 
\begin{equation}
\omega =\left\{ 
\begin{array}{l}
2\gamma \left( m+1-\mu \right) ,\;\sigma =-1 \\ 
2\gamma m,\;\sigma =1
\end{array}
\right. .  \label{abe24}
\end{equation}
These expressions are modifications of Eqs. (\ref{abe21}), (\ref{abe22}) for
the case $B<0$.

2. Consider the spinors $u_{l}\left( r\right) $ satisfying (\ref{abe26}).\
In the case $\omega =0$ they read 
\begin{equation}
u_{0,l}(r)=\left( 
\begin{array}{l}
\phi _{0,l,1}(r) \\ 
0
\end{array}
\right) ,\mathrm{\;}{}l\geq 1;\;\;\;u_{0}^{II}(r)=\left( 
\begin{array}{l}
\phi _{0,1}^{ir}(r) \\ 
0
\end{array}
\right) ,\;{}l=0.  \label{abe29}
\end{equation}
In the case $\omega \neq 0$ they are 
\begin{eqnarray}
&&u_{m,l,\pm }(r)=\left( 
\begin{array}{l}
\phi _{m,l,1}(r) \\ 
\mp i\phi _{m,l,-1}(r)
\end{array}
\right) ,\mathrm{\;}{}l\leq -1\;,{}\;\omega =2\gamma \left( m-l+1-\mu
\right) ,  \notag \\
&&u_{m+1,l,\pm }(r)=\left( 
\begin{array}{l}
\phi _{m+1,l,1}(r) \\ 
\pm i\phi _{m,l,-1}(r)
\end{array}
\right) ,\mathrm{\;}{}l\geq 1\;,{}\;\omega =2\gamma \left( m+1\right) , 
\notag \\
&&u_{m+1,\pm }^{II}(r)=\left( 
\begin{array}{l}
\phi _{m+1,1}^{ir}(r) \\ 
\pm i\phi _{m,0,-1}(r)
\end{array}
\right) ,\mathrm{\;}{}l=0\;,{}\;\omega =2\gamma \left( m+1\right) ,  \notag
\\
&&u_{m,\pm }^{I}(r)=\left( 
\begin{array}{l}
\phi _{m,0,1}(r) \\ 
\mp i\phi _{m,-1}^{ir}(r)
\end{array}
\right) ,\mathrm{\;}{}l=0\;,{}\;\omega =2\gamma \left( m+1-\mu \right) .
\label{abe30}
\end{eqnarray}
These expressions are modifications of Eqs. (\ref{abe27}), (\ref{abe28}) for
the case $B<0$.

3. In the case $\omega =0$ only positive energy solutions (particles) of Eq.
(\ref{abe12}) are possible. They coincide with the corresponding spinors $u$
up to a normalization constant: 
\begin{equation}
_{+}\psi _{0,l}(r)=Nu_{0,l}(r){},\;l\geq 1;\;\;\;_{+}\psi
_{0}^{II}(r)=Nu_{0}^{II}(r){},\;l=0.  \label{abe33}
\end{equation}
Thus, particles have the rest energy level, and the antiparticle states
spectrum begins from $_{-}\varepsilon =-\sqrt{M^{2}+2\gamma }$.

4. Relations for the irregular spinors $u_{\omega ,\sigma }(r)$, similar to
ones (\ref{abe35}), for the case $B<0$ have the form 
\begin{equation}
\Pi u_{\omega ,-1}(r)=i\sqrt{2\gamma }u_{\omega ,1}(r),\;\;\Pi u_{\omega
,1}(r)=-i\frac{\omega }{\sqrt{2\gamma }}u_{\omega ,-1}(r)\,.  \label{abe36}
\end{equation}

\end{document}